\documentclass[fleqn,usenatbib,useAMS]{mnras}

\usepackage[T1]{fontenc}

\usepackage{ae,aecompl}

\usepackage{graphicx}   

\usepackage{amsmath}    

\usepackage{amssymb}    

\newcommand{\Teff}{\ensuremath{T_{\rm eff}}}                

\newcommand{\logg}{\ensuremath{\log g}}                     

\newcommand{\Msun}{\ensuremath{\,{\rm M}_\odot}}            

\newcommand{\Rsun}{\ensuremath{\,{\rm R}_\odot}}            

\newcommand{\kms}{\,km\,s$^{-1}$}                           

\newcommand{\cmss}{\,cm\,s$^{-2}$}                          

\newcommand{\Veq}{\ensuremath{V_{\rm eq}}}                  

\newcommand{\Vsini}{\ensuremath{v \sin i}}                  

\newcommand{\Vsync}{\ensuremath{V_{\rm synch}}}             

\title[Conservative mass transfer in $\delta$~Lib]
{Evidence for conservative mass transfer in the classical Algol system $\delta$~Librae
from its surface carbon-to-nitrogen abundance ratio}

\author[A.~Dervi\c{s}o\u{g}lu et al.]
       {A.~Dervi\c{s}o\u{g}lu$^{1,2}$, K.~Pavlovski$^1$, H.~Lehmann$^3$,
         J.~Southworth$^4$ and  D.~Bewsher$^5$\\
  $^1$\,Department of Physics, Faculty of Science, University of Zagreb,
        Bijeni\v{c}ka cesta 32, 10000 Zagreb, Croatia \\
   $^2$\,Department of Astronomy \& Space Sciences, Erciyes University, Kayseri, Turkey \\
   $^3$\,Th\"{u}ringer Landessternwarte Tautenburg, Sternwarte 5, D-07778 Tautenburg, Germany \\
   $^4$\,Astrophysics Group, Keele University, Staffordshire, ST5 5BG, UK \\
   $^5$\,Jeremiah Horrocks Institute, University of Central Lancashire, Preston,
         Lancashire PR1 2HE, UK \\
}

\date{}

\pubyear{2018}

\begin{document}

\label{firstpage}

\pagerange{\pageref{firstpage}--\pageref{lastpage}}

\maketitle

\begin{abstract}

Algol-type binary systems are the product of rapid mass transfer between the 
initially more massive component to its companion. It is still unknown 
whether the process is conservative, or whether substantial mass is lost 
from the system. The history of a system prior to mass exchange is imprinted 
in the photospheric chemical composition, in particular in the 
carbon-to-nitrogen (C/N) ratio. We use this to trace the  efficiency of 
mass-transfer processes in the components of a classical Algol-type 
system, $\delta$~Librae. The present analysis is based on new spectroscopic 
data (ground-based high-resolution \'echelle spectra) and  extracted 
archival photometric 
observations (space-based measurements from the STEREO satellites).
In the orbital solution, non-Keplerian effects on the radial-velocity 
variations were taken into account. This reduces the primary's mass by 1.1 \Msun\ 
($\sim$23\%) significantly in comparison to previous studies, and removes a long-standing
 discrepancy between the radius and effective temperature.
A spectral disentangling technique is applied to the \'echelle observations 
and the spectra of the individual components are separated. Atmospheric 
and abundance analyses are performed for the mass-gaining component and we found C/N $= 1.55 \pm 0.40$ 
for this star. An extensive set of evolutionary models ($3.5\times10^6$) 
for both components are calculated from which the best-fitting model is 
derived. It is found that $\beta$, the parameter which quantifies the 
efficiency of mass-loss from a binary system, is close to zero. This means 
that the mass-transfer in $\delta$~Lib is mostly conservative with little 
mass loss from the system.
\end{abstract}


\section{Introduction}

\label{sec:intro}

Algol-type binary systems are comprised of components which are apparently in 
conflicting evolutionary states. The less massive star is a giant or subgiant,
 whilst the more massive component is a main-sequence star. A more precise 
definition is ``Algols are generally taken to be close, usually eclipsing, 
semi-detached binary systems consisting of an early type (A or late B, 
sometimes F) primary, which seems quite similar to a Main Sequence star 
in its broad characteristics, accompanied by a peculiar low-mass star, 
which fills, or tends to overflow, its surrounding `Roche' surface of 
limiting dynamical stability'' \citet{Budding_review}. The Algol paradox, 
as the contradictory evolutionary state of the components is colloquially 
called, was resolved with a plausible hypothesis of mass transfer from what 
was initially the more massive star in the system, in accordance with the 
theory of stellar evolution. Budding's definition states that the more 
massive component became dynamically unstable first, causing mass loss 
after filling its Roche lobe.

The mass transfer process happens on a thermal time scale, thus it is 
evolutionarily quite short, and the process is rapid. In turn, the mass ratio 
reversal between components happens, and further mass transfer occurs 
on a long, nuclear, time scale. In terms of the Roche gravitational 
equipotential, the binary system is in a semi-detached configuration as an 
Algol-type binary is defined \cite[c.f.][]{Hilditch}.

Large-scale mass transfer in Algol-type binary systems not only leads to 
mass reversal and an exchange of the role between the components but also 
affects the photospheric chemical composition. Formerly deep layers can 
become exposed at the surfaces after a short-lived phase of mass exchange. 
Therefore, tracing the photospheric elemental abundance pattern of the 
components might help to constrain their past, and in particular their 
initial stellar parameters. In particular, the carbon-to-nitrogen (C/N) 
ratio of the photospheric abundances for donor and gainer components in 
Algol-type systems is a sensitive probe of the thermohaline mixing 
(c.f.~\citet{Sarna}, and references therein). Constraining the mixing 
processes allows the determination of the initial binary and stellar 
parameters.

In turn, this opens the possibility to discriminate between conservative 
and non-conservative mass transfer in Algol-type systems, which is still 
one of the main unknown processes in understanding the evolution of binary 
stars \citep{Dervisoglu}.

A general trend which obeys theoretical predictions with nitrogen 
over-abundance and carbon depletion relative to solar abundance of 
these species has been detected in pioneering studies \citep{Plavec, 
Parthasarathy, Dobias, Cugier, Tomkin_five}. In addition, the equivalent 
widths (EWs) of the prominent resonant \ion{C}{ii} $\lambda$4267\,{\AA} 
line were measured for the mass-gaining stars in a large sample (18) of 
Algols and compared to normal B-type stars that have the same effective 
temperature (\Teff) and luminosity class \citep{Ibanoglu}.
The carbon under-abundance was clearly demonstrated, and was estimated to be 
on average $-0.54$ dex relative to the photospheric abundance of normal 
main-sequence B-type stars. The authors also demonstrated the relationship 
between the EWs of the \ion{C}{ii} $\lambda$4267\,{\AA} line and the rate 
of orbital period increase, i.e.\ the rate of mass transfer.

The cooler mass-losing component of the Algol-type system will have a small 
but non-negligible contribution to the line depths of the spectra of the 
system. This effect needs to be accounted for in spectral analysis, so that
 the parameters of the hotter mass-gaining component are accurate.

Individual spectra of the components of the binary system can be obtained 
by spectral disentangling of a time-series of observed spectra \citep{tomo91,
 spd94, had95}. Thus, disentangled components' spectra in turn make possible
 a precise determination of the atmospheric parameters, and chemical 
composition \citep{hensberge_2000, pav_2005}. The spectral disentangling 
technique has previously been used to constrain the initial parameters 
for the system and components of the hot Algol-type system u Her 
\citep{kolbas_uher} and to obtain a complete optical spectrum of 
the faint subgiant component in Algol itself \citep{kolbas_algol}.

Progress in understanding the mass and angular momentum transfer in 
interacting binary systems of Algol type is possible only with increased 
accuracy of stellar and system parameters, and constraining evolutionary 
models with the photospheric chemical composition. In this sense, of 
particular importance is an empirical determination of the carbon-to-nitrogen 
abundance ratio in the photospheres of the stars which experience the phase 
of rapid exchange of matter and angular momentum.

The structure of this paper is as follows. Section \ref{sec:overview} gives 
a brief description of $\delta$~Lib, and a short overview of previous 
research. In Sect.\,3 we describe our new high-resolution \'echelle 
spectroscopy, and STEREO photometry. Determination of new improved orbital 
elements with an extensive discussion  of non-Keplerian effects on the RV 
variations follows in Sect.\,4. An application of spectral disentangling 
yields separated spectra of the components used for an accurate determination
 of the atmospheric parameters for both components (Sect.\,5).
The STEREO light curve (LC) of $\delta$~Lib is analysed in Sect.\,6. 
The results of the abundance analysis for star A is given in Sect.\,7, 
whilst the evolutionary modelling is presented and discussed in Sect.\,8. 
In Sect.\,9 we summarise our findings for $\delta$~Lib.

\begin{figure*}
\includegraphics[width=15cm]{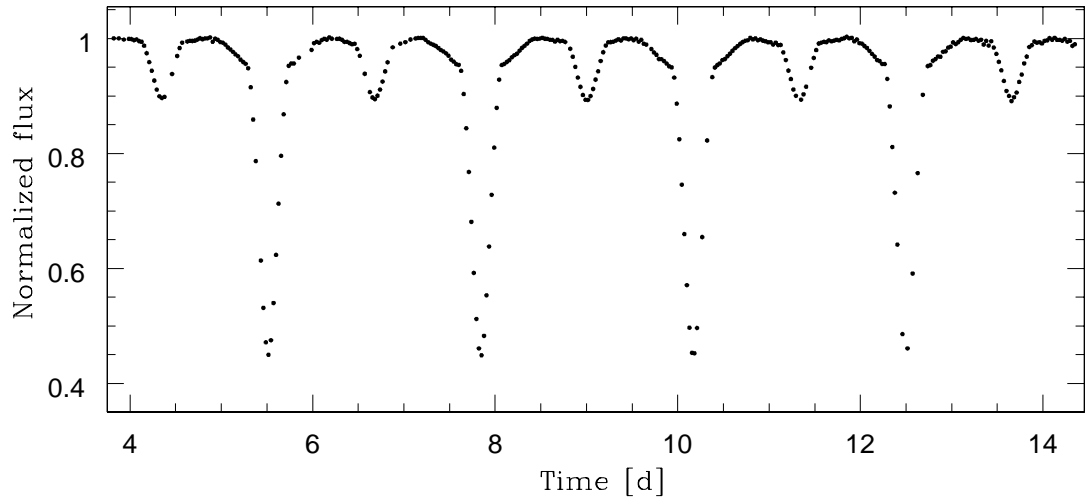}
\caption{Time-series of photometric measurements for $\delta$~Lib obtained 
with the Heliospheric Imager (HI-1) on-board the STEREO-A satellite covering 
five consecutive orbital cycles. HI-1 observations have a cadence of about 
40 minutes. These data are analysed in Section\,\ref{sec:lcsol}.}
\label{fig:stereo}
\end{figure*}

\section{Brief overview of $\delta$~Lib}

\label{sec:overview}

The bright eclipsing binary $\delta$~Lib is a classical Algol-type system. 
The hotter and more massive star is A0 and on the main sequence; we will 
refer to it a star A. Its companion, which we call star B, is a cooler and 
less massive K0 subgiant filling its Roche lobe. Mass is being transferred 
from star B (the mass-donor) to star A (the mass-gainer).

Despite being one of the brightest eclipsing binaries in the northern sky, 
$\delta$~Lib has not attracted much attention among photometrists. This is 
probaly because of the difficulty in finding appropriate comparison stars 
for such a bright object. The only complete broad-band LCs in the optical 
are due to \citet{Koch_UBV}. Koch observed $\delta$~Lib in 1956 and 1957, 
with a small additional set of observations secured in 1961. A photoelectric
 photometer, equipped with a filter set which resembled the $UBV$ photometric
 system, attached to the Steward Observatory 36-inch telescope was used. 
The comparison star used was about 1$^\circ$ from the variable. Koch was 
aware of the difficult observational circumstances, with a low position 
on the sky for a northern hemisphere observatory, and a fairly large 
angular distance between the comparison star, check star and $\delta$~Lib. 
Koch's photometric solution of the $UBV$ LCs was complemented by the RV 
measurements of \citet{Sahade_rvs}. Without a spectroscopic detection of 
the fainter companion, \cite{Sahade_rvs} estimated the components' masses 
using the mass function, $f(M)=0.115$\Msun, and an assumed value for the mass 
of star A. They recognised that star B is the less massive of the two components.

Between 1991 and 2001, \citet{Shobbrook} observed $\delta$~Lib using the 61\,cm 
telescope at Siding Spring Observatory. A photoelectric photometer with 
Str\"{o}mgren $b$ and $y$ filters was used. Targets, including $\delta$~Lib, 
were observed in the $y$ passband, whilst the $b$ passband was used only for 
observing standard stars, and the transformation of $y$ passband measurements 
to the Johnson $V$ scale. \citet{Shobbrook} published the LC of $\delta$~Lib 
along with {\it Hipparcos} measurements which were secured on-board the satellite 
from 1989 to 1993. Transforming both the \citet{Shobbrook} and {\it Hipparcos} 
measurements to Johnson $V$ produces a consisent dataset. The phase coverage is 
good, but sparse.

\citet{Lazaro_infrared} presented the first infrared LCs of $\delta$~Lib in the 
$JHK$ passbands. Their observations were obtained with the 1.5-m Carlos Sanchez 
Telescope at the Observatorio del Teide, Tenerife between 1994 and 1998, with 
a mono-channel photometer and a cooled InSb detector. The authors analysed their 
LCs and those of \citet{Koch_UBV}. The spectroscopic mass ratio was available for 
their LC analysis.

They calculated the \Teff\ of the primary component from the calibration of the 
Str\"{o}mgren colour indices using measurements of $\delta$~Lib close to the 
secondary minimum obtained by \citet{Hilditch_Hill}, and the $\beta$ index from 
\citet{Hauck_Mermilliod} from the observations out of eclipse. They suggested 
for the primary star a \Teff\ in the range 9650 to 10\,500\,K and a surface gravity, 
\logg, in the range 3.75--4.0. They found a considerable discrepancy of the components' 
properties (radii, luminosities) versus their spectroscopic masses. Hence, they 
considered `a low-mass' model adopting a mass of $M_1 \sim 2.85$\Msun\ to reconcile 
the over-sized and over-luminous components with their dynamical masses.

The faintness of evolved companions in Algol-type binary systems is a principal 
observational obstacle in studies of these (post) mass-transfer systems. Progress 
was very slow until the mid-1970s when the first efficient red-sensitive electronic 
detectors became available. In the first paper of his breakthrough series on 
discoveries of spectral lines of faint secondaries in Algol-type binaries, 
\citet{Tomkin_delta} detected and measured radial velocities (RVs) of the 
\ion{Ca}{ii} near-infrared triplet at 8498--8662\,\AA. From dynamics, 
\citet{Tomkin_delta} determined the mass ratio, $q = 0.347\pm0.043$, with 
the individual masses $M_1 = 4.9\pm0.2$\Msun, and $M_2 = 1.7\pm0.2$\Msun, 
for the (hotter) primary and (cooler) secondary, respectively.

The mass ratio determined by \citet{Tomkin_delta}, and questioned by the LC analysis 
of \citet{Lazaro_infrared}, was corroborated with new spectroscopic observations 
by \citet{Bakis}. Spectra of a short wavelength interval centred on H$\alpha$ were 
secured with the Ond\v{r}ejov and Ro\v{z}hen observatories 2\,m telescopes, 
respectively, in 1996-1997, and 2003. Using a method of spectral disentangling 
they obtained the RV semi-amplitudes, $K_1 = 81.8\pm4.8$\kms\ and $K_2 = 
213.8\pm4.8$\kms, and the mass ratio, $q = 0.383\pm0.058$. In turn, this gave 
masses close to the previous result of Tomkin, $M_1 = 4.7\pm0.3$\Msun, and 
$M_2 = 1.8\pm0.3$\Msun. The analysis of \citet{Bakis} is mainly focused on 
the presence of a third star. In a comprehensive analysis of the {\it Hipparcos} 
astrometry the authors quantified the characteristics of its orbit and its physical 
properties.

\begin{figure*}

\includegraphics[width=18cm]{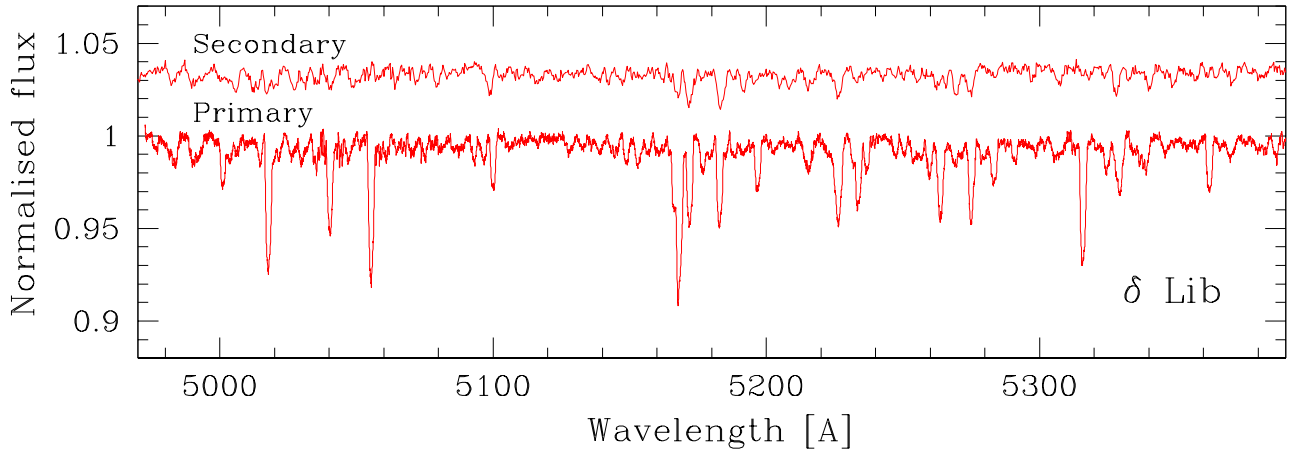}

\caption{Disentangled spectra for the primary and secondary components in 
$\delta$~Lib. Metal lines in both spectra are weak but with the light ratio 
$l_{\rm sec}/l_{\rm prim} = 0.053$ in the $V$ passband this is intrinsic for 
the primary's spectrum. The weakness of the secondary's disentangled spectrum 
is due to its small fractional light contribution, $l_{\rm sec}$, to the total 
light of the binary system.}

\label{fig:disent}

\end{figure*}

The relative radius of star A (the radius of the star in units of the binary separation)
 is a large fraction of its Roche lobe, and an accretion disc cannot be formed around 
this component according to the \citet{lubow} criterion. No emission in H$\alpha$ was 
detected in the spectroscopic survey of \citet{Richards_halpha}. An extensive Doppler 
tomographic reconstruction of $\delta$~Lib \citep{Richards_tomo} has revealed several 
components indicating activity and interaction: (i) some weak emission which is 
characteristic of most Algol-type systems, and is associated with the activity of 
a cool mass-losing component, (ii) indications of a gas stream along the predicted 
gravitational trajectory between the two components,  and (iii) a bulge of extended 
absorption/emission around the mass-gaining component produced by the impact of 
the high-velocity stream on to the relatively slowly rotating photosphere. 
Compared to other Algols, $\delta$~Lib is a less `active' system, as was revealed 
in a comprehensive tomographic study by \citet{Richards_tomo}.

With a distance of about 95\,pc, and with a cool giant secondary component, 
$\delta$~Lib is a bright source in the radio sky and was detected early in radio 
surveys \citep{woodsworth, radio, micro}. The analysis of the continuous and 
long-term radio monitoring of radio flares from $\delta$~Lib confirmed its 
low activity, known from the optical spectral range \citep{Richards_radio}. 
No periodicity was determined from occurrences of radio flares for $\delta$~Lib 
in contrast to $\beta$~Per for which some strong radio flares were detected, 
and the periodicity of the flares appears to be $\sim$50\,d. If $\delta$~Lib 
were at the distance of $\beta$~Per, the radio flux of $\delta$~Lib would be
 much smaller than $\beta$~Per. \citet{Richards_radio} explained the low level 
of radio flux and lower flare activity of $\delta$~Lib as being due to the higher
 \Teff\ of star B, G1\,IV compared to K2\,IV for $\beta$~Per. The possibility still 
exists that \citet{Richards_radio} observed $\delta$~Lib in a quiescent period since 
it was found by \citet{radio} as variable radio source, whilst \citet{singh_xray} 
detected significant intensity variations in X-ray emission.

\section{The photometric and spectroscopic observations}

\label{sec:newobs}

\subsection{STEREO photometry}

\label{sec:stereo}

The NASA Solar TErrestrial RElations Observatory ({\sc stereo}) was launched 
in October 2006 and is comprised of two nearly identical satellites 
\citep{stereo}. One is ahead of the Earth in its orbit (STEREO-A), 
the other trailing behind (STEREO-B). With their unique view of the 
Sun-Earth line, the STEREO mission has revealed the 3D structure of 
coronal mass ejections (CMEs), and traced solar activity from the Sun 
and its effect on the Earth. One of the principal instruments on board 
is the Heliospheric Imagers (HI-1) \citep{helio_imagers}. These have 
a large field of view, 20$\times$20$^{\circ}$, and are centred 14$^{\circ}$ 
away from the limb of the Sun. The stability of the instruments has allowed 
high-quality photometry of background point sources.

Over the course of an orbit almost 900\,000 stars brighter than about 12 mag 
are imaged within 10$^{\circ}$ of the ecliptic plane. Each STEREO satellite 
observes the stars for about 20 days with gaps of about one year. HI-1 
observations have a cadence of about 40 minutes which make them very 
suitable for studies of eclipsing binary systems. \citet{Wraight_stereo} 
used STEREO/HI observations to detect 263 eclipsing binaries, about half 
of which were new detections. The LCs can be affected by solar activity. 
Another problem is severe blending in a crowded stellar field due to the 
low image resolution of about 70 arcsec.

The STEREO observations of $\delta$~Lib presented in \citet{Wraight_stereo} 
were obtained between July 2007 and Aug 2010, with a total time span of 1113 
days. The quality of the photometric 
observations from HI-1A are of higher 
quality than from HI-1B, due to some systematic effects. In total 5895 
measurements were obtained,  of which 2802 with HI-1A imager.
 Due to enhanced solar coronal activity, we 
selected a time span  the observations were clean from CMEs. The 
STEREO photometric measurements of $\delta$~Lib used here cover a time 
span of 11.80 days beginning with MJD 24554377.54433.  In total 362  ($\sim$13\%) 
measurements  from the HI-1A imager  smoothly covering five complete orbital 
cycles of the binary (Fig.\,\ref{fig:stereo}) were adopted for the analysis
in this work. These measurements needed only a small correction for
detrending, and evenly cover the phased light curve. 
The CCDs on board the HI instruments have 2048$\times$2048 
pixels, but they are binned on board to 1024$\times$1024 pixels. The 
pointing information in the fits headers (which were calibrated by 
\citealt{Brown_stereo}) was used to locate the star in an image, and 
then perform aperture photometry on that position. This process was 
repeated for all images that contain the star.

\begin{table}
\caption{RV measurements for our high-resolution spectra.
BJD is barycentric Julian Day of midexposure. The phases are calculated 
with the \citet{Koch_UBV} ephemeris. Subscripted 1 denotes star A, and 
subscripted 2 is for star B. The observatories are abbreviated as follows: 
AS for Asiago Astrophysical Observatory, CA for Calar Alto, TU for TUG 
for T\"{U}B\.{I}TAK National Observatory, and TL for Th\"{u}\-rin\-ger 
Landes\-stern\-warte Tau\-ten\-burg.}
\begin{tabular}{ccrrrrl}
\hline

BJD	& Phase &  RV$_1$    & $\sigma_1$	&  RV$_2$	& $\sigma_2$    & Obs. \\

\hline

54907.55260	&  0.243	&  -122.9	&  5.0	&  165.0   &  5.0	& AS	\\

54908.52634	&  0.661	&  22.8	&  5.0	&  -233.0	 &  5.0	& AS	\\

54988.34355	&  0.957	&  -7.0	&  3.1	&   	-        &  - 	&  CA	\\

54988.34796	&  0.959	&  0.3	&  1.4	&	-        &  -	        &  CA	\\

54988.35570	&  0.962	&  8.6	&  2.4	&	-        &  -	        &  CA	\\

54988.49871	&  0.023	&  -84.8	&  3.1	&	-       &   -	&  CA \\

54989.45560	&  0.435	&  -69.0	&  1.4	&  73.0	 &  10.0	&  CA	\\

54989.46330	&  0.438	&  -67.2	&  1.1	&  69.6  	 &  7.0	&  CA	\\

54989.47762	&  0.444	&  -65.5	&  3.5	&  67.0	 &  4.0	&  CA	\\

54989.51552	&  0.460	&  -57.5	&  2.5	&  69.9	 &  7.0	&  CA	\\

54989.52205	&  0.463	&  -56.2	&  1.6	&  60.4	 &  6.0	&  CA	\\

54989.52855	&  0.466	&  -55.1	&  2.2	&  65.3	 &  6.5	&  CA	\\

54990.38109	&  0.832	&  27.3	&  1.7	&  -205.6	&  4.2	&  CA	\\

54990.38659	&  0.835	&  24.9	&  1.4	&  -209.7	&  1.3	&  CA	\\

54990.38986	&  0.836	&  25.3	&  1.0	&  -215.0	&  1.6	&  CA	\\

54990.42792	&  0.852	&  22.2	&  2.0	&  -190.0	&  2.5	&  CA	\\

54990.43094	&  0.857	&  20.4	&  2.8	&  -203.0	&  2.0	&  CA	\\

54990.43398	&  0.855	&  21.1	&  1.9	&  -192.0	&  2.4	&  CA	\\

54990.52673	&  0.895	&  7.1	&  3.1	&  -162.2	&  2.3	&  CA	\\

54990.52991	&  0.896	&  4.4	&  2.1	&  -173.5	&  4.8	&  CA	\\

54990.53300	&  0.897	&   4.8	&  3.7	&  -154.0	&  2.9	&  CA	\\

54991.54431	&  0.332	&  -110.4	&  1.2	&  136.7	&  2.7	&  CA	\\

54991.54737	&  0.333	&  -109.0	&  1.6	&  139.4	&  2.0	&  CA	\\

54991.55274	&  0.336	&  -108.0	&  1.9	&  151.4	&  2.4	&  CA	\\

54991.55814	&  0.338	&  -106.5	&  1.9	&  133.4	&  2.1	&  CA	\\

55346.30550	&  0.763	&  39.0	&  5.0	&  -233.0	&  5.0	&  TU	\\

57493.50854	&  0.359	&  -100.1	&  2.3	&  136.6	&  5.0	&  TL	\\

57496.51161	&  0.649	&  24.0	&  1.1	&  -229.4	&  2.8	&  TL	\\

57498.47693	&  0.494	&  -40.2	&  0.4	&     -	& 	-       &  TL	\\

57499.52097	&  0.942	&  -4.2	&  1.8	&  -74.7	&  5.0	&  TL	\\

57499.54255	&  0.952	&  0.9	&  3.1	&  -62.2	&  3.3	&  TL	\\

57500.46108	&  0.346	&  -103.7	&  2.9	&  135.6	&  6.7	&  TL	\\

57502.53648	&  0.238	&  -116.7	&  2.8	&  168.7	&  4.3	&  TL	\\

57507.52080	&  0.380	&  -93.0	&  0.6	&  125.2	&  5.8	&  TL	\\

\hline

\end{tabular}

\label{tab:rvs}

\end{table}

\subsection{High-resolution \'echelle spectroscopy}

Spectroscopic observations were secured at two observing sites. A set of 
23 spectra was obtained in two observing runs (May and August 2008) at 
the Centro Astron\'{o}nomico Hispano Alem\'{a}n (CAHA) at Calar Alto, 
Spain. The 2.2\,m telescope equipped with the {\sc foces} \'{e}chelle 
spectrograph \citep{foces} was used. The spectra cover 3700--9200\,{\AA} 
at a resolving power of $R \approx 40\,000$. {\sc foces} used a Loral\#11i 
CCD detector. To decrease the readout time, the spectra were binned 
2$\times$2. The wavelength calibration was performed using a thorium-argon 
lamp, and flat-fields were obtained using a tungsten lamp.
The observing conditions were generally good in both observing runs.

The spectra were bias subtracted, flat-fielded and extracted with 
{\sc iraf}\footnote{{\sc iraf} is distributed by the National Optical 
Astronomy Observatory, which are operated by the Association of the 
Universities for Research in Astronomy, Inc., under cooperative agreement 
with the NSF.} \'{e}chelle package routines. Normalisation and merging 
of the \'{e}chelle orders was performed with great care, using programs 
described in \citet{kolbas_algol}, to ensure that these steps did not 
cause systematic errors in the reduced spectra.

A set of eight spectra was secured with the Coud\'e \'{e}chelle spectrograph 
attached to the 2\,m Alfred Jensch Telescope at the Th\"{u}ri\-nger 
Landes\-stern\-warte Tau\-ten\-burg. They cover 4540--7540\,{\AA} at 
$R \approx 30\,000$. The spectrum reduction was performed using standard 
ESO-MIDAS packages. It included filtering of cosmic ray events, bias and 
straylight subtraction, optimum order extraction, flat fielding using 
a halogen lamp, normalisation to the local continuum, wavelength calibration
 using a thorium-argon lamp, and merging of the \'{e}chelle orders. 
The spectra were corrected for small instrumental shifts using a large 
number of telluric O$_2$ lines.

\section{Determination of the orbital elements}

\label{sec:orbit}

The faintness of star B relative to star A makes RV measurements difficult. 
\citet{Tomkin_delta} presented the first detection of star B's spectrum in 
$\delta$~Lib, and the only other attempt to isolate the secondary component 
was presented by \citet{Bakis}. \citet{Tomkin_delta} measured the RVs of both 
components from the \ion{Ca}{ii} triplet at 8498--8662\,{\AA}. He detected 
star B's spectrum in 18 of 21 available spectra obtained with the Reticon 
spectrograph at the 2.7\,m telescope McDonald Observatory in 1977. Spectral 
disentangling \citep{spd94, had95}, herewith {\sc spd}, was applied by 
\citet{Bakis} to derive the orbit for both components of $\delta$~Lib.

\citet{Bakis}  used new sets of eight and nine spectra obtained on 
2\,m telescopes and covering a small spectral interval (6300--6750\,{\AA}) 
centred on H$\alpha$.

Our new observations, comprising 31 \'echelle spectra, were first 
disentangled in Fourier space \citep{had95} using the {\sc FDBinary} code 
\citep{sasa_fdb}. Spectral disentangling clearly resolved the spectra of
both components in spite the secondary component contributes much less than
the primary component in the optical part of the spectrum ({Fig.\,\ref{fig:disent}). 
This preliminary analysis returned the RV semiamplitudes  
$K_{\rm A} = 78.9\pm 1.2$\kms\ and $K_{\rm B} = 201\pm 1.8$\kms, giving a mass ratio 
of $q = 0.391\pm0.007$.  

The {\sc iSpec} software \citep{refispec} was used to measure RV 
shifts of lines with the cross-correlation method. {\sc iSpec} is able to 
measure cross-correlation functions with a built-in spectra synthesizer 
for very broad spectral ranges. We used the atmospheric parameters of 
each component from Section \ref{sec:atmos}, to measure each component 
separately because of a large difference between the \Teff s of the 
components.

We also used several archival spectra of $\delta$~Lib from the Asiago 
and TUG observatories \citep{Ibanoglu}.

For the cool secondary component, only a red part of the spectra was used 
since its fractional light contribution is increasing towards the red, 
and is about 13\% in the spectral interval 6000--7000\,{\AA}.

The RV measurements are given in Table \ref{tab:rvs}. Utilising the initial 
solution from {\sc spd}, we analysed the $O-C$ residuals of the FOCES and 
TLS RVs separately in order to derive a possible systematic difference of 
the systemic velocity ($\gamma$) from two datasets. We found that the two 
$\gamma$s agree to within 2$\sigma$: $\gamma_{\rm CAHA} = -40.52\pm 0.37$\kms\
 and $\gamma_{\rm TLS} = -39.65 \pm 0.40$\kms. An offset of $0.87 \pm 
0.54$\kms\ in the RVs between both data sets could be due to the zero-point 
difference between the two spectrographs, or to the influence of a third 
body in the system.

The presence of the third body was claimed by \citet{Worek} from his 
comprehensive study of all available (historic) century-long RVs measurements 
for the primary component of $\delta$~Lib, including new observations. 
He derived a possible orbit of the tertiary component with the period 
$P = 2.762$ yr, and the semi-amplitude of the changes in $\gamma$ 
velocities of ~5\kms. Our new determinations of the $\gamma$ velocity 
for the {\sc CAHA} and {\sc TLS} data sets are shown in Fig.\,\ref{fig:worek}.
 Also, in Fig.\,\ref{fig:worek}, we show the results for the $\gamma$ velocity
 determined in \citet{Bakis} Fig.\,\ref{fig:worek} shows that the modern 
values of the $\gamma$ velocity contradict \citeauthor{Worek}'s solution 
based on historic photographic spectral plates with mostly low resolution. 
Our measurements corroborate the conclusion from \citet{Bakis} that currently 
the detection of the third component in the $\delta$~Lib system is not 
supported by spectroscopic observations.

\begin{figure}
\includegraphics[width=8.5cm]{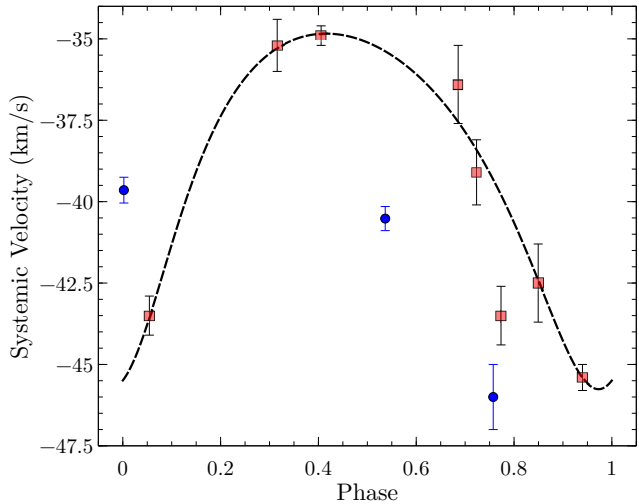}
\caption{The variations in a systemic velocity $\gamma$. The solid line 
represents orbital variations due to the third body as determined by 
\citet{Worek} from a collection of historic data and including his new 
observations. Data used by \citet{Worek} are represented with red squares. 
Modern determinations of $\gamma$ from \citet{Bakis} and the current work 
are shown as blue circles, and do not corroborate the suggestion that an 
outer orbit with the period $P \approx 2.7$ yr exists.}

\label{fig:worek}

\end{figure}

In Algol-type binary systems, the cool component is filling its critical 
equipotential surface, i.e.\ in terms of the Roche equipotential surface, 
the system is in a semi-detached configuration. However, a tidal distortion 
of the Roche-lobe filling component causes non-Keplerian effects on the RV 
curve. The proximity distorts the component(s) from a spherically symmetric 
shape, and the optical centres of the stars depart from the mass centres. 
The elaborate analytical derivation of these tidal and rotational distortions 
were performed by \citet{Kopal1980a, Kopal1980b}. \citet{Wilson79} implemented 
a computer algorithm to calculate these effects in the widely-used {\sc wd} 
code \citep{wd} based on the theoretical development of \citet{wilson-sofia}.

Moreover, a recent analysis of \citet{Sybilski} based on 12\,000 simulations 
of binary populations shows the importance of tidal effects on the orbital 
parameters. They introduce the parameter
$s = \frac{R_1+R_2}{a(1-e)}$
as an indicator of tidal effects and departures from Keplerian orbits. They 
conclude that for $s \geq 0.5$, the tidal distortion introduces a deviation 
up to 10\,km/s on the RVs semi-amplitudes and up to 0.01 pseudo-eccentricity.

To account for the non-Keplerian effects, we use the {\sc wd} code through 
the {\sc phoebe}  \citep{phoebe} implementation to model the RV measurements.
 We implemented 
the Markov chain Monte Carlo (herein {\sc mcmc}, see \citet{sharma}, and 
refences therein) to derive the orbital elements and their uncertainties.

As expected, the proximity effects on the RV semi-amplitude of star B are 
prominent, while they are negligible for the tidally undistorted star A. 
The difference in the RV semi-amplitude 
of star B between the pure Keplerian and non-Keplerian (tidally distorted) 
solutions amounts to about 7\kms. When translated into the mass of star A, 
the difference is about 0.35\Msun\ (c.f.\ Table\,\ref{tab:orbit}).

An examination of the residuals for the Roche-lobe filling star B indicates 
that both models (pure Keplerian, and tidally distorted non-Keplerian) do 
not fully explain the RV distortions during both ingress and egress of 
occultations. It looks like the optical centre of star B is shifted toward 
the Lagrangian L2 point. Such distortion was also noticed in \citet{Bakis}. 
In fact, \citet{Budding} have explained the distortions in the line profiles 
of close binary systems with the presence of circumstellar matter, which is 
not unexpected in Algol-type systems. As already mentioned in 
Sect.\,\ref{sec:overview} tomographic reconstruction for $\delta$~Lib by 
\citet{Richards_tomo} revealed several structures in this particular 
location between the components. In the last attempt to account for 
distortions in star B's RV curve we modelled the system with an artificial 
spot on the star, mimicking obscuring material around the L1 point in 
between the stars. The final fit is shown in Fig.\,\ref{fig:rv} with no 
systematic deviations present in the residuals of either component. 
An obscuration (spot) has a strong impact on the RV semi-amplitude of 
star B, which is now $K_{\rm B} = 192.0\pm2.9$\kms\ (c.f.\ Table\,\ref{tab:orbit}),
or about 20\kms\ and 30\kms\ less than in the Keplerian and tidally 
deformed non-Keplerian solutions, respectively. 
It is well documented in the literature that spot solutions in the light
curves always bear 
some degree of degeneracy due to an extended parameter space (cf.~\cite{Morales}).
In our {\sc mcmc} calculation of spot solution, 
we have found a strong correlation between spot radius and the RVs 
semiamplitude $K_{\rm B}$ as is obvious from Fig.\,\ref{fig:mcmc_spot}. 
This corelation is also projected itself to the error estimations. 
The uncertainity on 
$K_{\rm B}$ from spot solution is nearly four times bigger than those 
from Keplerian and tidal one (Table\,\ref{tab:orbit}). Translated to masses, 
this is a drop of about 0.9\Msun\ for star A compared to the pure Keplerian 
solution, or about 1.3\Msun\ compared to the tidally deformed non-Keplerian 
solution. Since, star A's RV semi-amplitude is barely affected, changes in 
the mass for star B are small, about 0.22 and about 0.3 \Msun\ for both 
solutions, respectively. The geometry of the system and position of the
obscured material (`spot') on mass-loosing component are shown in
Fig.\,\ref{fig:roche}.

A similar effect has been detected for RZ~Cas, another short-period Algol 
system, by \citet{Tkachenko_rzcas}. They discussed extensively the possible 
effects that might change the distribution of flux over the stellar disc 
and thus affect the RV curve. The solution could be improved taking into 
account the influence of a disc of variable density and a dark spot on 
the cool component.

\begin{table}

\caption{The orbital solutions for $\delta$~Lib: (a) solution for a pure 
Keplerian orbit, (b) tidal distortion of the Roche-lobe filling star B 
taking into account in the RV variations, and (c) the influence of dark 
obscuring material around the Lagrangian L1 point on the RV variations 
of star B also taking into account (a) and (b). In the calculations this 
obscuring cloud is represented as a dark `spot' located on star B, in the 
direction facing star A (c.f.~Fig.\,\ref{fig:roche}).}

\begin{tabular}{lccc}

\hline

Param./Orb.     &   Keplerian	&   Tidal  & Dark `spot'    \\

\hline

K$_{\rm A}$ [\kms]	 & $79.8 \pm 0.5  $ & $ 79.4 \pm 0.4  $ &  $ 79.9 \pm 0.5$  \\

K$_{\rm B}$ [\kms]	 & $211.9 \pm 0.8  $ & $219.2 \pm 0.7  $ &  $192.0 \pm 2.9$  \\

$\gamma$ [\kms]	 & $-40.7 \pm 0.1  $ & $-39.8 \pm 0.1  $ &  $-41.1 \pm 0.1$  \\

$q$                & $0.377{\pm}0.003$ & $0.362{\pm}0.002$ &  $0.416{\pm}0.007$  \\

\hline

Spot (donor)   & & & \\

\hline

 Colat.\ [rad]      &   &   & $0.906{\pm}0.006$     \\

 Long.\ [rad]        &   &   & $ 6.06 \pm 0.04$    \\

 Rad.\ [rad]        &   &   & $ 1.23 \pm 0.08$    \\

$T_{\rm spot}/T_{\rm B}$  &   &   & $ 0.34 \pm 0.09$    \\

\hline

$M_{\rm A} \sin i$ [\Msun]   & $ 4.35 \pm 0.04$  &  $ 4.71 \pm 0.04$  &  $ 3.42 \pm 0.12$   \\

$M_{\rm B} \sin i$ [\Msun]   & $ 1.64 \pm 0.02$  &  $ 1.71 \pm 0.02$  &  $ 1.42 \pm 0.03$   \\

$a \sin i $  [\Rsun]   & $13.41 \pm 0.04$  &  $13.73 \pm 0.04$  &  $12.50 \pm 0.13$   \\

\hline

\end{tabular}

\label{tab:orbit}

\end{table}

\begin{figure*}

\includegraphics[width=14cm]{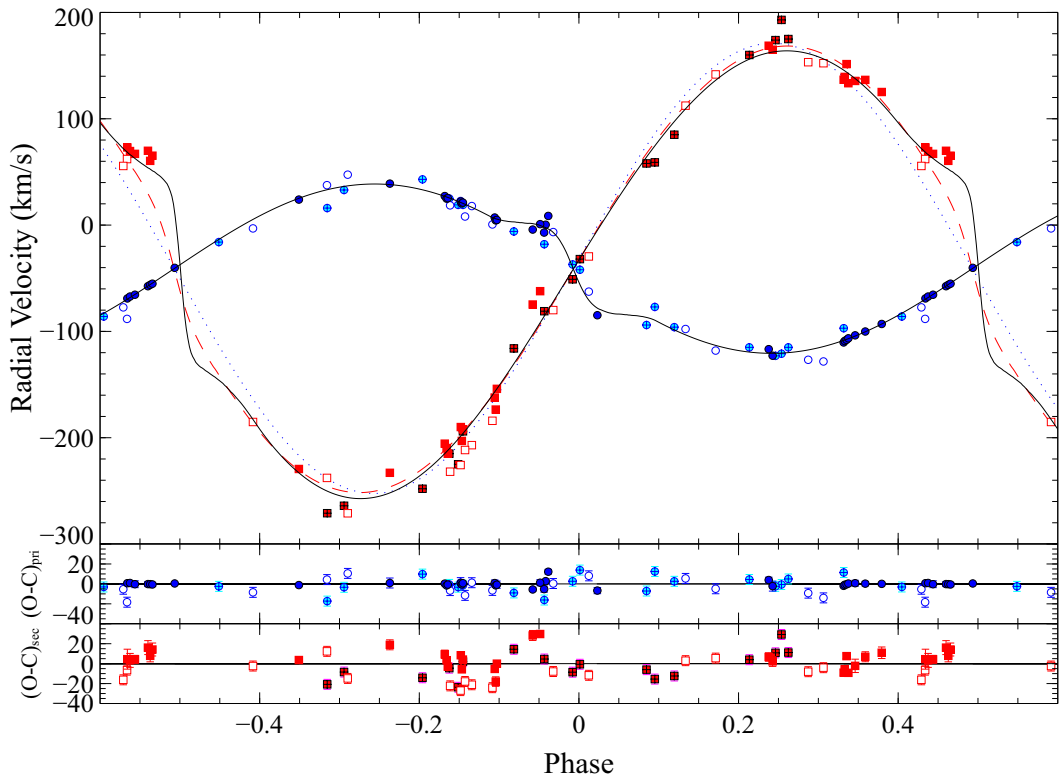}

\caption{The best fit (solid curve) to measured RVs of star A (blue circles),
and star B (red squares), with crossed symbols for \citet{Tomkin_delta}, 
open symbols for
\citet{Bakis}, and solid symbols for our RVs measurements. 
In the solution non-Keplerian effects are taken into 
account: the RM effect during eclipses, tidal distortion of star B, and dark 
obscuring material between the components, around the Lagrangian L1 point. 
The solutions for a pure Keplerian orbit (dotted curve), and with tidal 
effects on both stars (dashed curve) are also shown for comparison.}

\label{fig:rv}

\end{figure*}

\begin{figure}
\includegraphics[width=8.5cm]{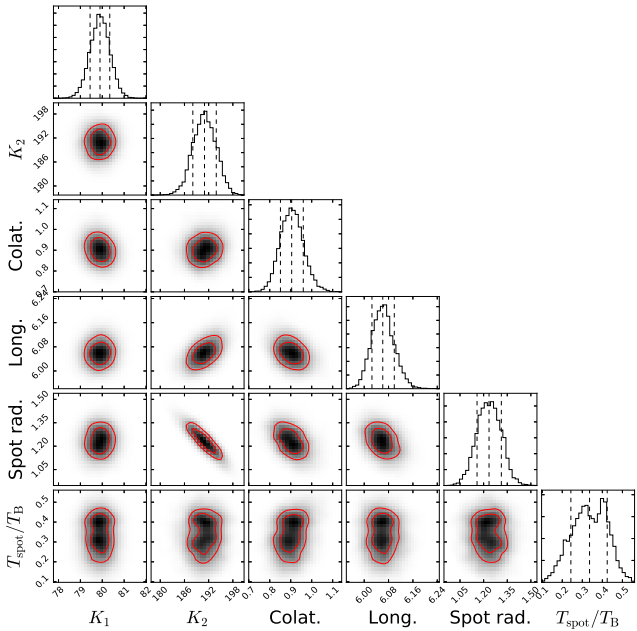}
\caption{ The determination of the orbital and spot parameters and accompaning
uncertainties from measured RVs using {\sc phoebe} code. 
Distribution and correlation of these parameters from 
the $10^6$ {\sc mcmc} runs are shown.  Degeneracy in determination of the
temperature ratio $T_{\rm spot}/T_{\rm B}$ is clearly seen. Also, a strong
correlation between spot radius and the RV semiamplitude of star B (mass-loosing
component) is present. 
In each section, the
map of distribution densities between parameters are plotted with
associated 1\,$\sigma$  and 2\,$\sigma$  confidence levels (solid countours).
 The histogram distributions  (solid
lines) are plotted across the associated variable. Mean and 1 $\sigma$ of each
variable are indicated with dashed lines.}

\label{fig:mcmc_spot}

\end{figure}

\begin{figure}
\includegraphics[width=8cm]{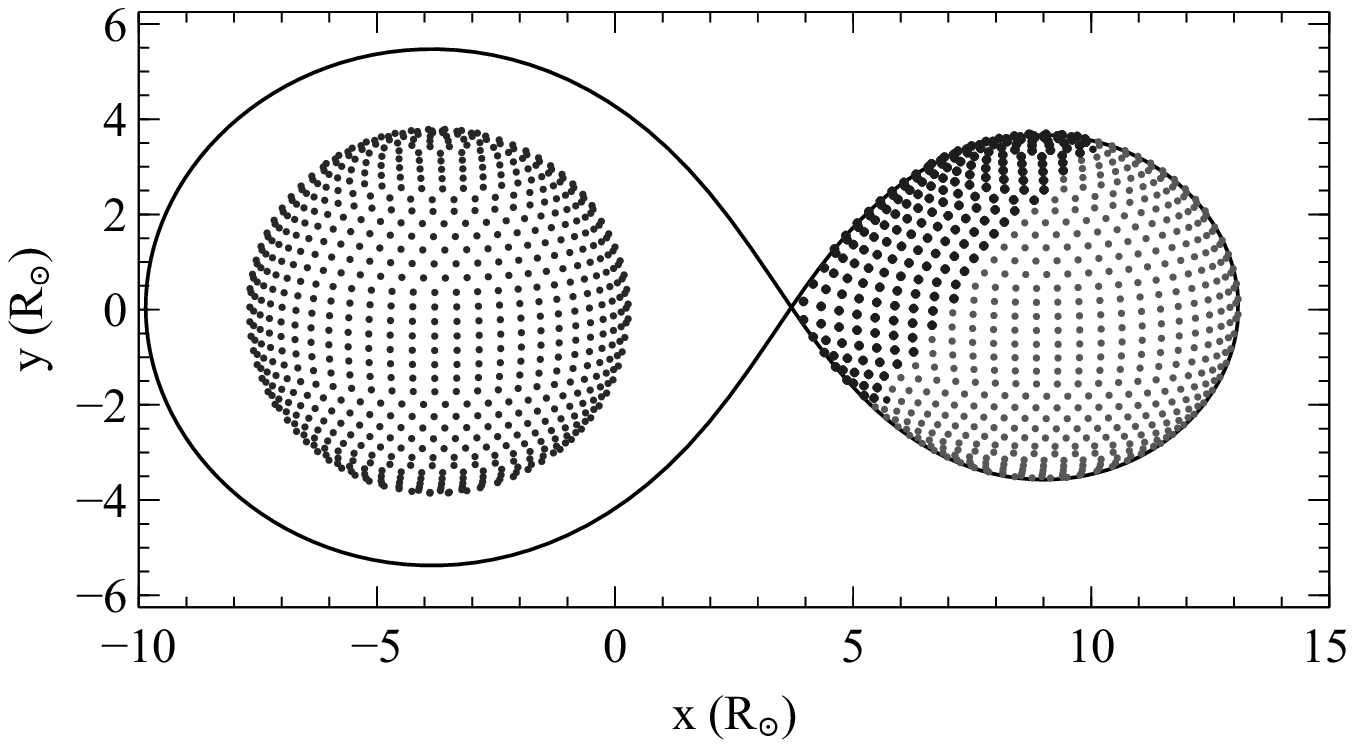}
\caption{Roche geometry for the $\delta$~Lib system. The secondary, less
massive and cooler component is filling its Roche lobe, whilst the primary
component is detached from its Roche lobe. The position of the obscuring
material in the vicinity of the L1 point, modelled as a dark spot on the
secondary component, is indicated.}

\label{fig:roche}

\end{figure}

\section{Atmospheric parameters for the components}  \label{sec:atmos}

As described in the preceding section, due to the  non-Keplerian effects 
on the RV variations of the components in the $\delta$~Lib system, plus 
the prominent Rossiter-McLaughlin (RM) effect, i.e.~line profiles distortions
 in the course of the eclipses \citep{Rossiter, McLaughlin},  spectral 
disentangling in its complete form, i.e.\ determination of the orbital 
elements, and reconstruction of the individual spectra of the components, 
cannot be applied. Therefore, spectral disentangling was performed in a pure 
separation mode \cite[c.f.][]{pav_brno} using the RVs measured by 
cross-correlation (Table\,\ref{tab:rvs}), and {\sc cres} code \citep{cres} 
based on the {\sc spd} method of \citet{spd94}. Although the spectra are 
separated, the continuum is still composed of the light from both stars. 
To determine the atmospheric parameters the disentangled spectra should be 
renormalised to their own continuum. One possibility is to use the light 
ratio between the components from the LC analysis 
\cite[c.f.][]{hensberge_2000}. The other is a direct optimal fitting of 
disentangled spectra in which the light ratio is to be determined too. 
Information on the light ratio is preserved in disentangled spectra, 
as well as projected rotational velocity \citep{tamajo_genfit}.

 In the present analysis, we used both options and obtained very consistent 
results. The {\sc starfit} code \citep{kolbas_uher} is used for the optimal fitting 
of the disentangled spectrum of star A. Star B contributes little to the 
total light of binary system, about 5--9 per cent in the optical part, and 
with a rather limited number of spectra available for {\sc spd} its 
disentangled spectrum suffers from a low S/N.

 Three hydrogen lines exist in the spectral range of the disentangled 
spectrum of star A: H$\beta$, H$\gamma$, and H$\delta$ (Fig.\,\ref{fig:balmer}). 
The {\sc starfit} 
code searches a large 2D grid (\Teff\ v.\ \logg) of pre-calculated 
theoretical spectra, and compares them to the disentangled spectrum. 
The following parameters can be obtained: \Teff, surface gravity \logg, 
light dilution factor $ldf$, projected rotational velocity \Vsini, 
relative velocity shift between the disentangled spectrum and the 
laboratory rest-frame wavelengths of the theoretical spectra $v_0$, 
and the continuum offset correction $ccor$. For the optimisation of 
these parameters a genetic algorithm \citep{pikaia} is adopted. 
The grid of theoretical spectra are calculated in LTE with the program 
{\sc uclsyn} \citep{smith, smalley}.

In the course of the present study, a Monte Carlo Markov Chain (MCMC) 
algorithm for the calculations of the uncertainties in the optimal fitting 
of the disentangled spectra has been implemented in the {\sc starfit} code. 
The most general MCMC algorithm is the Metropolis-Hasting (MH) algorithm 
\citep{Metropolis, Hastings}
which was used for the error analysis (c.f.~\citet{Hogg}, 
and references therein). For the 
simulation, we used at least $10^5$ steps after the burn-in phase. 
To ensure good mixing in the Markov chain, we used an adaptive step 
in our algorithm and ensured our acceptance ratio was bigger than 
0.279 as suggested by \citet{sharma}.

\begin{table*} 
\centering
\caption{Atmospheric parameters for $\delta$~Lib\,A determined from 
optimal fitting of disentangled spectra centred on the Balmer lines.}

\label{tab:mcmc}

\begin{tabular}{lcccc}

\hline

Spectral line: H$\beta$                 & Unit    & Optimisation 1  & Optimisation 2 & Optimisation 3    \\

\hline

Effective temperature, \Teff\           &  K      & $10562 \pm 60   $ & $10560 \pm 60$    & $10519 \pm 30.4$   \\

Surface gravity, $\log g$               &  [cgs]  & $3.835 \pm 0.008$ &  3.84 fixed       & 3.84 fixed  \\

Light dilution factor, $ldf$             &         & $0.938 \pm 0.009$ & $0.937 \pm 0.009$ & Normalised  \\  

Projected rotational velocity, $v\sin i$ &  km/s   & $72.52 \pm 3.18 $ & $   72 \pm  3   $ & 73 fixed  \\

\hline

Spctral line: H$\beta$ Degenerate solution  &     &   &  &     \\

\hline

Effective temperature, \Teff\           &  K      & $ 9914 \pm 54   $   & --  & --    \\

Surface gravity, $\log g$               &  [cgs]  & $3.573 \pm 0.005$   & --  & --    \\

Light dilution factor, $ldf$             &         & $0.918 \pm 0.008$   & --  & --    \\

Projected rotational velocity, $v\sin i$ &  km/s   & $73.80 \pm 2.77 $   & --  & --    \\

\hline

Spectral line: H$\gamma$                &     &   &  &     \\

\hline

Effective temperature, \Teff\           &  K      & $10756 \pm 158  $ &  $10770 \pm 60$    & $10550 \pm 66.1$         \\

Surface gravity, $\log g$               &  [cgs]  & $ 3.82 \pm 0.04 $ &  3.84 fixed        & 3.84 fixed    \\

Light dilution factor, $ldf$             &         & $0.984 \pm 0.021$ &  $0.984 \pm 0.009$ & Normalised    \\

Projected rotational velocity, $v\sin i$ &  km/s   & $ 73.4 \pm 6.5  $ &  $73 \pm 6$        & 73 fixed    \\

\hline

Spectral line: H$\delta$                &     &   &  &      \\

\hline

Effective temperature, \Teff\           &  K      & $10724 \pm 149  $ &  $10610 \pm 150$    & $10484 \pm 82$         \\

Surface gravity, $\log g$               &  [cgs]  & $ 3.94 \pm 0.03 $ &  3.84 fixed         & 3.84 fixed    \\

Light dilution factor, $ldf$             &         & $0.967 \pm 0.022$ &  $0.965 \pm 0.025$  & Normalised   \\

Projected rotational velocity, $v\sin i$ &  km/s   & $ 82.0 \pm 12.7 $ &  $82 \pm 10 $       & 73 fixed    \\

\hline

Adopted weighted $T_{\rm eff}$    &  K      &   &      & $10520 \pm 110$        \\

\hline

\end{tabular}
\\
\end{table*}

As mentioned above \citet{Lazaro_infrared} determined the \Teff\ and 
\logg\ for star A from photometric indices. Various calibrations have 
restricted \Teff\ to between 9650 and 10\,500\,K, and \logg\ to between 
3.75 and 4.0. In this \Teff\ range the degeneracy between \Teff\ and 
\logg\ is present for the strong and broad Balmer lines. However, in 
eclipsing and double-lined spectroscopic binaries the masses and radii 
could be determined with high accuracy, and the degeneracy between \Teff\ 
and \logg\ in the wings of the Balmer lines can be lifted with a precisely 
determined \logg. The LC analysis also provides the light ratio between 
the components thus making determination of the \Teff\ from disentangled 
spectra more certain. In this work, we performed an optimal fitting of 
the disentangled primary's spectrum using three methods: (1) all parameters 
free, (2) \logg\ fixed, and (3) \logg\ and light dilution factor fixed. 
These parameters are listed in Tables \ref{tab:lcsol} and \ref{tab:abspar}. 
Only a few iterations were needed to converge to the final values. The 
results of all three fits are given in Table\,\ref{tab:mcmc}.

As expected, the solutions for option (1) are degenerate. 
In Table\,\ref{tab:mcmc} both solutions are given for the H$\beta$ line. 
This is also illustrated in Fig.\,\ref{fig:starfit}  in which two solutions 
are clearly distinguishable, one at $\Teff = 9910$\,K and $\log g = 3.57$, 
and the other at $\Teff = 10\,560$\,K and $\logg = 3.84$. It is interesting 
that in the second solution the \logg\ matches perfectly the value determined 
from the primary's dynamical mass and radius from the LC solution. Without 
the possibility to obtain the \logg\ from combined RVs and LC analysis, the 
atmospheric parameters could be erroneous due to a strong degeneracy between 
the \Teff\ and $\log g$ in the Balmer lines for hot stars. It is also 
encouraging that fixing $\log g$ gives a light dilution factor in perfect 
concordance with the LC solution (optimisation 2 in Table\,\ref{tab:mcmc}).

Having lifted the degeneracy between \Teff\ and $\log g$, and with an 
appropriate renormalisation of the disentangled spectra using the light 
ratio from the photometric analysis, the uncertainties in the determination 
of $\Teff$ have been reduced. This is important, first, for the quality 
of the LC solution, and, second, for proper calculations of the model 
atmospheres and the determination of photospheric abundances. Since 
the LC solution and the \Teff\ determination from the disentangled 
spectra are interconnected, an iterative approach is needed. 
Convergence was fast, and only a couple of iterations were 
needed for consistent solutions.  The quality of the fits for
the Balmer lines, H$\delta$, H$\gamma$, and H$\beta$ are shown in
Fig.\,\ref{fig:balmer}.

Projected rotational velocities, $v\sin i$, for both components were 
determined with {\sc starfit} by fitting metal lines. The Balmer lines 
and severely blended metal lines were masked.  Such fitting gives 
$v\sin i = 72.5\pm3.2$\kms. Two previous results for star A's \Vsini\ 
were published:  \citet{Cugier} determined $v\sin i = 75$\kms\ (with 
no uncertainty), whilst \citet{Parthasarathy} found a slightly lower 
value of $v\sin i = 69\pm4$\kms. Our determination is within 1$\sigma$ 
of both values.

The large uncertainty in determination of star B's \Vsini\ does not allow 
any further discussion on its implication for constraining the mass ratio 
through rotational synchronisation of this Roche-lobe filling component.

In the renormalisation of the disentangled spectra for star B, the 
multiplicative factor is large (about 11 and 20 in $V$ and $B$ passbands, 
respectively) due to its small contribution to the total light of the 
system. Accordingly the noise is multiplied, and determination of the 
atmospheric parameters is less reliable than from the LC analysis 
(Tables\,\ref{tab:lcsol} and \ref{tab:abspar}). Still, we are interested 
to see if the optimal fitting would give a sensible solution. 
The {\sc starfit} solution with the MCMC error calculations gives 
the following atmospheric parameters for the faint secondary component: 
$\Teff = 5095\pm280$ K, $\logg = 3.48\pm0.07$, $ldf = 0.052\pm0.009$, 
and $v \sin i = 94\pm12$\kms. This should be compared with the values 
we adopted in the present work from the RVs and LC modelling: $\Teff 
= 5134\pm240$ K, $\logg = 3.48\pm0.06$, $ldf = 0.068\pm0.015$ 
(interpolated between $B$ and $V$ passbands in Table\,\ref{tab:lcsol}).

\begin{figure}
\includegraphics[width=\columnwidth]{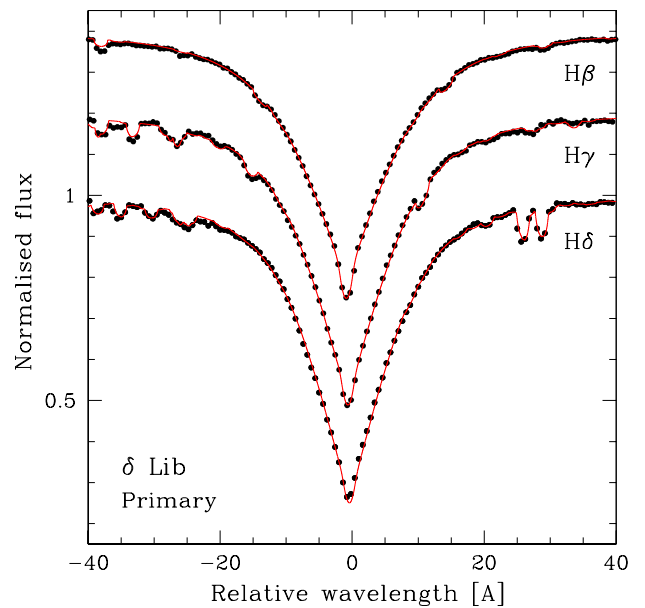}
\caption{The best-fitting synthetic spectra (lines) compared to the
renormalised disentangled spectra (filled circles) of star A. H$\beta$
(upper), H$\gamma$ and H$\delta$(lower) profiles are shown with offsets
for clarity.}

\label{fig:balmer}

\end{figure}

\begin{figure}
\includegraphics[width=8.2cm]{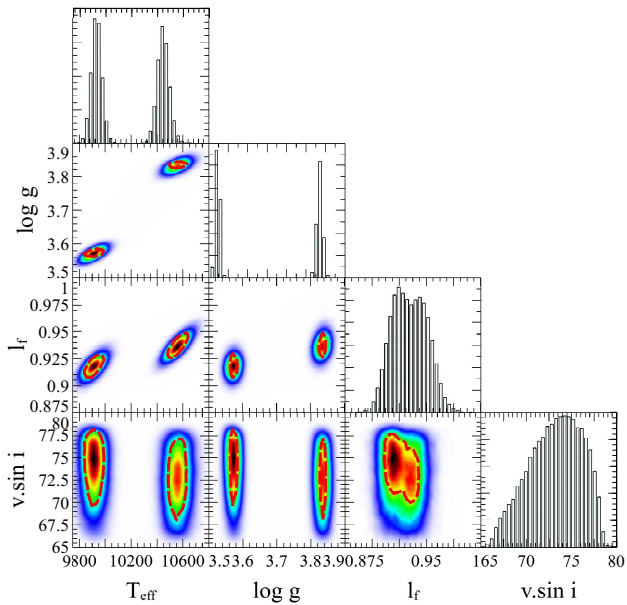}
\caption{The determination of atmospheric parameters with  MCMC through
10$^6$ steps from the H$\beta$ line. Degeneracy in the solution is obvious,
and external information is needed to break it. We used precise determination
of \logg\ from the combined RV and LC analysis which then yield star A's
\Teff. Different options in an optimal fitting of the Balmer lines are given
in Table\,\ref{tab:mcmc}.}

\label{fig:starfit}

\end{figure}

\section{Light curve analysis}   \label{sec:lcsol}

The following photometric observations of $\delta$~Lib are used in the 
light curve analysis (see Section\,\ref{sec:overview}):

\begin{itemize}

\item Broadband photoelectric photometry of \citet{Koch_UBV}. These total 
337 measurements in $U$, 338 in $B$, and 343 in $V$.

\item Str\"{o}mgren $y$ photometry from \citet{Shobbrook}, which has been 
transformed to the Johnson $V$ passband. The dataset consists of 111 
measurements that are fairly well distributed in orbital phase.

\item The {\it Hipparcos} data comprise 89 measurements in the original 
{\it Hipparcos} passband and transformed to Johnson $V$ \citep{harmanec1998}.

\item Infrared $JHK$ photometry of \citet{Lazaro_infrared}. The majority 
of the observations are in the $JHK$ filters with a few measurements in $L$.
 In total, the photometry consists of 722 measurements in $J$, 783 in $H$, 
and 770 in $K$.

\item Broadband photometry extracted from the STEREO/ He\-lio\-sphe\-ric Imagers 
images \citep{helio_imagers}. The data acquisition is described in 
Sect.\,\ref{sec:stereo}.

\end{itemize}

We used the {\sc phoebe} suite 0.31a \citep{phoebe}, which is based on work by 
\citet{wd} and generalized by \citet{Wilson79}, to find a solution to 
the LCs. The mass ratio $q = 0.416 \pm 0.007$ is determined in 
Sect.\,\ref{sec:orbit}, and was fixed in the analysis of the LCs. 
Since the $\delta$~Lib system is in a semi-detached configuration, 
the radius of star B is defined by the mass ratio and the geometry 
of the system. Thus {\sc mode 5} (semi-detached binary with the 
secondary component filling its Roche lobe) was used in the calculations.  
The \Teff\ for star A is determined from the optimal fitting of its 
disentangled spectrum (Sect.\,\ref{sec:atmos}), In the final runs we
 fixed it to $T_{\rm eff,\,A} = 10\,520$ K. Gravity-darkening coefficients 
were set to $g_{\rm A} = 1.0$ and $g_{\rm B} = 0.32$ which is appropriate for 
radiative \citep{vonZeipel}, and convective envelopes \citep{lucy1967}, 
respectively. In accordance with the properties of the atmospheres, 
we kept fixed the bolometric albedos $A_{\rm A} = 1$ and $A_{\rm B} = 0.5$ 
after \citet{Rucinski}. {\sc phoebe}'s implementation of logarithmic limb 
darkening interpolation uses \cite{vanhamme} tables, with the exception 
of the $H$ passband which was taken from \citet{Claret_limb}.

The following parameters were left free for adjustment in the LC calculations:
 the \Teff\ of star B ($T_{\rm eff,B}$), the surface potential of star A 
$\Omega_{\rm A}$ (i.e.\ its relative radius $r_{\rm A}$), and the orbital 
inclination $i$. Table\,\ref{fixedpars} provides the list of fixed parameters
 during this analysis. The orbital phases are calculated with the ephemeris 
given by \citet{Koch_UBV}. The best fit and the uncertainties for adjusted 
parameters are then calculated with {\sc phoebe} with the {\sc mcmc} implementation.

\begin{table} \centering
\caption{The fixed system parameters of $\delta$~Lib during the lightcurve 
analysis.}

\label{fixedpars}

\begin{tabular}{llr}

\hline

Parameter                         & Unit   & Value               \\

\hline

Orbital period $P$                & d      &  $2.32735297$       \\

Primary eclipse time HJD          & d      &  $2\,422\,852.3598$ \\

Mass ratio $q$                    &        &  $0.416 \pm 0.007 $ \\

$T_{\rm eff}$ of star A           & K      &  $10\,520 \pm 110 $ \\

Gravity darkening  (A,B)     &        &  $1.0$, $0.32$       \\

Bolometric albedo  (A,B)     &        &  $1.0$, $0.5$        \\

Third light                       &        &  $0.0$               \\ [3pt]

\hline

\end{tabular}

\end{table}

\begin{figure*}
\includegraphics[width=15cm]{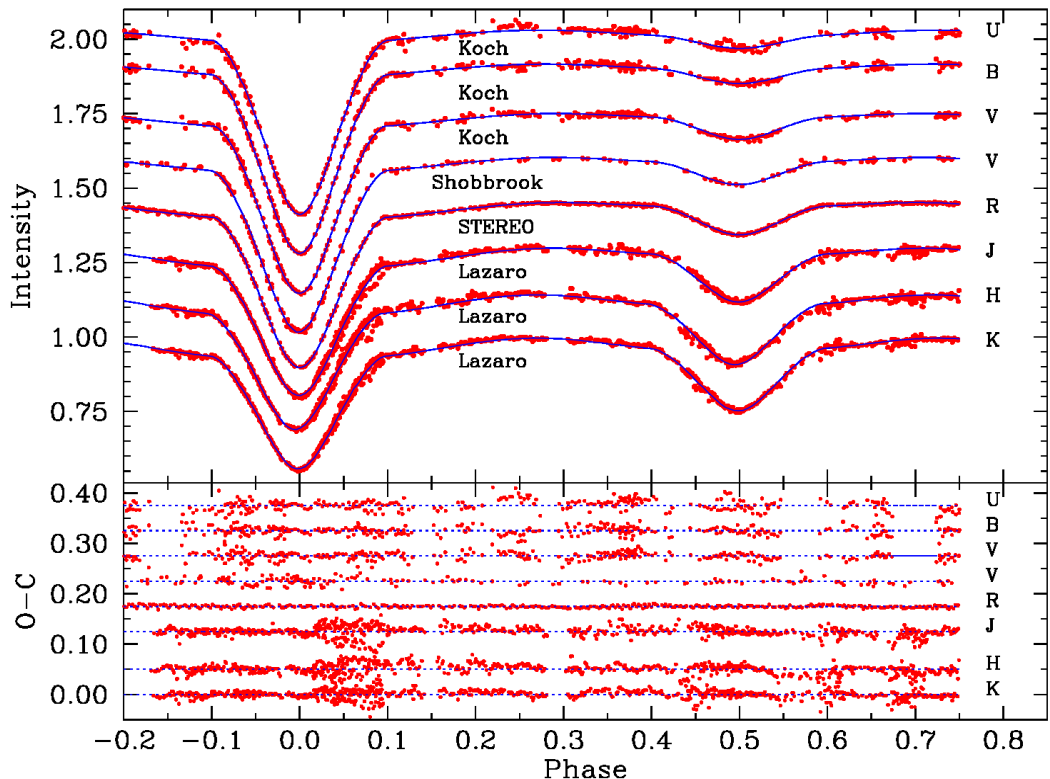}
\caption{Observed phased LCs of $\delta$~Lib (solid dots) with the 
best-fitting {\sc phoebe} model LCs (continuous line) with associated 
passband and observer labels (top panel). In the lower panel the residuals 
have been plotted to show the quality of the fit. Both lightcurves and 
residuals are shifted for the ease of viewing. The phases are calculated 
using the \citet{Koch_UBV} ephemeris. A small adjustment of the phases 
was needed, and was included in the LC analysis.}

\label{figLC}

\end{figure*}

For this purpose we wrote a home-made script for {\sc phoebe} to produce 
the {\sc mcmc} simulations. This script uses the Metropolis-Hastings 
algorithm to accept or reject the proposed 
value of the orbital parameters to find the most probable fitting model 
and associated errors. To calculate the acceptance ratios, we used the 
observational error (scatter of residuals) of the best-fitting parameter 
from the differential corrections algorithm in the WD code.

With starting random initial parameters and $10^5$ simulations for each 
LC after the burn-in phase, the uncertainties $\sigma$ for each free 
parameter are calculated from the normal distribution of accepted results.

Altogether, eight LCs are solved using the described procedure ($UBV$ from 
\citet{Koch_UBV}, $V$ from \citet{Shobbrook} and {\em Hipparcos}, {\sc stereo} 
LC in the $R$ 
band from the present study (see also \citet{Wraight_stereo}, and the 
$JHK$ data from \citet{Lazaro_infrared}.) The LCs, best fitting models 
and  residuals are shown in Fig.\,\ref{figLC}. The optimal parameters 
from {\sc phoebe} (as shown in Fig.\,\ref{figLC}) and the uncertainties 
from the {\sc mcmc} calculations  are given in Table\,\ref{tab:lcsol}.

\begin{table} \centering
\caption{Results from the solution of the {\sc wd} modelling, {\sc mcmc} 
analysis with associated $\sigma$ values for the free parameters, namely 
inclination, the \Teff\ of star B and the radius of star A. The light 
fraction and its errorbar for star A ($l_f=L_{\rm A}/(L_{\rm A}+L_{\rm B})$) and the scatter 
of the residuals of the fits are also listed.}

\label{tab:lcsol}

\begin{tabular}{cccccc}

\hline

Band	& $r_{\rm A}$ & $i$ [deg] & $T_{\rm B}$ [K] &  $l_{\rm A}$  & $\sigma$ [flux] \\

\hline

$U$   & 0.300	   & 80.6	  & 5468	& 0.9610       & 0.0300 \\

      & $\pm$0.005   & $\pm$0.25    & $\pm$182    & $\pm$0.0004  &       \\

$B$   & 0.300	   & 80.5	  & 5416	& 0.9520       & 0.0152 \\

      & $\pm$0.003  & $\pm$0.1    & $\pm$86     & $\pm$0.0005  &       \\

$V$   & 0.293	   & 80.1	  & 5261	& 0.9112       & 0.0150 \\

      & $\pm$0.003  & $\pm$0.1    & $\pm$60     & $\pm$0.0010  &       \\

$H_p$ & 0.301	   & 79.8	  & 5309	& 0.9120       & 0.0130 \\

      & $\pm$0.003  & $\pm$0.13    & $\pm$94    & $\pm$0.0011  &       \\

$R$   & 0.296	   & 79.22	  & 5126	& 0.8760       & 0.0048 \\

      & $\pm$0.001  & $\pm$0.05    & $\pm$17    & $\pm$0.0006  &       \\

$J$   & 0.296	   & 79.17	  & 5092	& 0.7574       & 0.0196 \\

      & $\pm$0.004  & $\pm$0.11    & $\pm$31     & $\pm$0.0052  &       \\

$H$   & 0.293	   & 79.33	  & 4994	& 0.6660       & 0.0220 \\

      & $\pm$0.005  & $\pm$0.12    & $\pm$44    & $\pm$0.0114  &       \\

$K$   & 0.299	   & 79.15	  & 5204      	& 0.6544       & 0.0176 \\

      & $\pm$0.004  & $\pm$0.11    & $\pm$40     & $\pm$0.0084  &       \\

\hline

Mean            & 0.297      & 79.45        &  5134        &   -          &    -   \\

$\sigma$     & $\pm$0.010 & $\pm$0.37    &  $\pm$ 240   &   -          &    -   \\

\hline

\end{tabular}

\end{table}




\begin{table} \centering \caption{The absolute dimensions and related 
quantities determined for $\delta$~Lib. \Vsync\ is the calculated synchronous 
rotational velocity.}

\begin{tabular}{llcc} \hline

 Parameter     & Unit      &      Star A    &      Star B    \\

\hline

Mass           &\Msun     & $   3.60 \pm 0.13$   &   $  1.50 \pm 0.04$    \\

Radius         & \Rsun     & $   3.78 \pm 0.13$   &   $  3.79 \pm 0.04$    \\

\logg\         & \cmss     & $   3.84 \pm 0.03$   &   $  3.46 \pm 0.01$    \\

\Teff          & K         & $10\,520 \pm 110 $   &   $5\,150 \pm 175 $    \\

$\log L$       & L$_\odot$ & $   2.19 \pm 0.03$   &   $  0.96 \pm 0.06$    \\

\Veq\ $\sin i$ &\kms      & $   72.5 \pm 3.2 $   &   $    94 \pm 12  $    \\

\Vsync\        &\kms      & $   82.1 \pm 2.9 $   &   $  82.3 \pm 1.0 $    \\

\hline

\label{tab:abspar}

\end{tabular}

\end{table}

\section{Abundance Analysis}

\label{sec:abund}

$\delta$~Lib\,A contributes the highest proportion of light to the total
 light of the binary system. Renormalisation of its disentangled spectrum degrades the S/N 
only slightly, which is not the case for $\delta$~Lib\,B. Thus, determination of the 
photospheric abundances was possible only for star A. The atmospheric parameters 
(\Teff\ and \logg), needed for the set-up of the model atmosphere are determined 
from an optimal fitting of star A's renormalised disentangled spectrum with fixed 
$\log g$, and calculated from its dynamical mass and radius determined from the LC
 analysis (Sections \ref{sec:orbit} and \ref{sec:lcsol}).

Due to the star A's relatively high \Vsini\ of $72.5\pm3.2$\kms, the spectral lines are 
broadened and blended. Only a few spectral lines, mostly of \ion{Fe}{ii}, are free of 
contamination, but their numbers are too low to determine the microturbulence velocity 
$\xi$ from the EWs. Instead, the elemental abundances were determined by an optimal
 fitting of selected line blends, for a fixed $\xi = 1.5$\kms. We adopted this value from 
the calibration work of \citet{gebran}. We selected line blends which are not too crowded, 
and which contain the CNO species of particular interest.

We used the {\sc uclsyn} code \citep{smalley} which allows a simultaneous fit for the five 
different atoms or ions. For such complex blends we iterated several times, usually starting 
iterations with different initial compositions. The line list and atomic data are compiled 
from the VALD database  \citep{vald1, vald2}.

In Table\,\ref{tab:abuall}, derived elemental abundances for star A are given. The uncertainties 
are calculated from a scatter (standard deviation) in the measurements for different lines and 
from differences in the abundances due to the uncertainties in the \Teff\ ($\pm110$\,K) and 
$\log g$ ($\pm0.023$). Also, we have taken into account the uncertainty in $\xi$ which we 
estimated as $\pm0.10$\kms.

Since the {\sc uclsyn} code  gives the rms values for each line, we calculated a grid of models 
within an error of the atmospheric parameters given above. We then used these solutions to find 
the scatter around our most probable solution. The uncertainties in star A's atmospheric parameters 
have a small contribution to the uncertainties in abundances which come mostly from the scatter 
from the spectral lines used in the determination. The abundances relative to solar composition 
given by \citet{asplund} are also given for comparison.

It turns out that the iron abundance [Fe/H] $=0.16 \pm 0.10$ is larger than the solar abundance.
An average photospheric abundance for metals (Mg, Si, Ti, Cr, and Fe) is [M/H] $= 0.16 \pm 0.13$, 
and supports the case for a higher metallicity determined for iron alone. Unfortunately in previous 
research the iron abundance was not determined and comparison is not possible. \citet{Cugier} 
determined the carbon abundance from \ion{C}{ii} resonant lines in the ultraviolet spectral region 
at 1335--1337\,{\AA} from the IUE satellite spectra. He found values of $\log \epsilon {\rm (C)} 
= 8.42\pm0.15$ in LTE, and $\log \epsilon {\rm (C)} = 8.36\pm0.20$ and $\log \epsilon {\rm (C)} 
= 8.45\pm0.20$ in NLTE, for the complete and partial redistribution, respectively. All Cugier's 
determinations are for these atmospheric parameters: $\Teff = 9900 \pm 200$\,K, $\log g = 4.00\pm0.10$,
 and $\xi = 2$\kms, and are determined from the ultraviolet flux distribution. Our determination is 
in almost perfect agreement with the carbon abundance determined by \citet{Cugier}, and also supports
 his conclusion of a solar standard value which is in modern evaluation $\log \epsilon {\rm (C)_\odot}
 = 8.43\pm0.05$ \citep{asplund}. The carbon abundance for $\delta$~Lib\,A was also estimated by 
\citet{Parthasarathy}. Their work was based on the analysis of the \ion{C}{ii} 4247\,{\AA} line. 
They found it rather weak in the spectrum, and gave only an upper limit of $\log \epsilon {\rm (C)} 
\leq 8.52\pm0.15$. This upper limit is within the uncertainty of 1$\sigma$ for both \citet{Cugier} 
and our measurements.

The enrichments of nitrogen and oxygen are obvious from our measurements,  with $\log \epsilon 
{\rm (N)} = 8.23\pm0.06$, and $\log \epsilon {\rm (O)} = 9.03\pm0.06$. The carbon-to-nitrogen 
abundance ratio is particularly important for our evolutionary modelling, which are given from 
our abundance determination (Table\,\ref{tab:abuall}) as $\log$ C/N $= 0.19 \pm 0.09$, i.e.\ 
C/N $= 1.55 \pm 0.40$.

\begin{table}

\centering \caption{\label{tab:abuall} Photospheric abundances
derived for $\delta$~Lib\,A. Abundances are expressed relative to the 
abundance of hydrogen, $\log \epsilon(H) = 12.0$. The third column gives 
the number of lines used. The fifth column lists the solar abundances 
from \citet{asplund} which are used as reference values.}

\begin{tabular}{lccccc} \hline

El 	& A & $N_{\rm li}$ &   $\log \epsilon({\rm X})$ 	&  Solar    & [X/H]	\\

\hline

C	&  6	&   6	&  8.42$\pm$ 0.11  & 8.43 $\pm$ 0.05   & -0.01 $\pm$0.09    \\

N	&  7	&   5	&  8.23	$\pm$ 0.06  & 7.83 $\pm$ 0.05  &  0.40 $\pm$0.06    \\

O	&  8	&  15	&  9.03	$\pm$ 0.06  & 8.69 $\pm$ 0.05  &  0.34 $\pm$0.06    \\

Mg	& 12	&  11	&  7.76	$\pm$ 0.17  & 7.60 $\pm$ 0.04  &  0.16 $\pm$0.12    \\

Si	& 14	&  7	&  7.67	$\pm$ 0.19  & 7.51 $\pm$ 0.03  &  0.16 $\pm$0.14    \\

Ti	& 22	&  23	&  5.21	$\pm$ 0.22  & 4.95 $\pm$ 0.05  &  0.26 $\pm$0.16    \\

Cr	& 24	&  15	&  5.70	$\pm$ 0.13  & 5.64 $\pm$ 0.04  &  0.06 $\pm$0.10    \\

Fe	& 26	& 146	&  7.65	$\pm$ 0.13  & 7.50 $\pm$ 0.04  &  0.15 $\pm$0.10    \\

\hline

\end{tabular}

\end{table}

\section{Evolutionary Analyses}

\label{sec:models}

\begin{figure}

\includegraphics[width=8.2cm, clip]{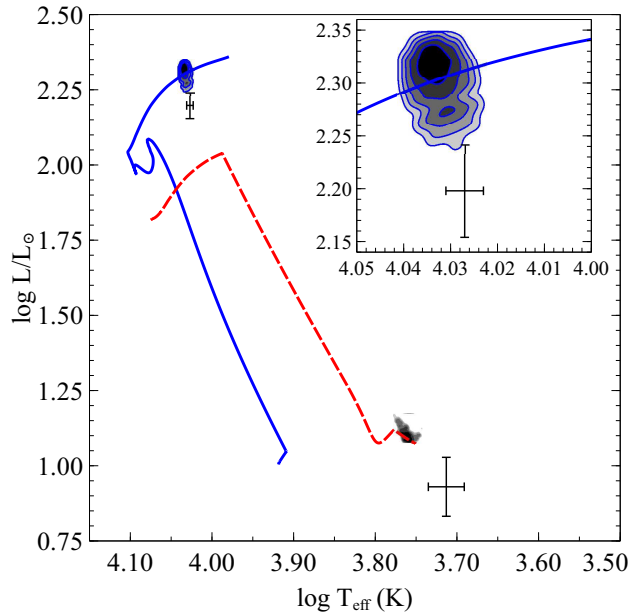}

\caption{The best fitting evolutionary tracks for the $\delta$~Lib mass-gainer (solid) and 
mass-loser (dashed) associated with empirical values from Table\,\ref{tab:abspar}. Density 
contours of all acceptable solutions are overplotted. The region around the mass-gainer is 
shown zoomed-in for clarity.}

\label{fighr}

\end{figure}

Evolutionary calculations that includes binary interactions (such as mass transfer, tidal 
sysnchronisations) have been done since mid-sixties (e.g. \cite{Kippenhahn1967}, 
\cite{Plavec1970}, \cite{Plavec1973}, \cite{Paczynski1971}). Currently, there are numerious 
groups conducted with 
different stellar evolution codes to model stellar response to the abrupt mass change 
triggered by Roche Lobe overflow (\cite{rensbergen}, \cite{Stancliffe}, \cite{siess}, 
\cite{paxton}) which most of them give similar results for accretion on non-degenerate stars. 
In general, the main uncertinities faced in modelling binary evolution is to reckoning the 
efficiency of mass transfer (i.e the mass loss from system) and the angular momentum loss. 
The problem is relatively easy for Algol-type systems since they are supposedly passed only 
one RLOF induced mass transfer. However, even with the mass-angular momentum conservation 
assumtion, one needs to take the initial mass ratio of system as free parameter. The best 
way to overcome such problem is to build sets of models with various systemic initial mass 
ratios and compare the evolution tracks with observations. The number of proposed models 
increased with power of free parameters when we also innclude the mass and angular momentum loss.

In the evolutionary analyses of binary stars, two different methodologies are often applied. 
The first one uses a large grid of binary evolution models which are calculated for a range 
of initial components' masses and periods to find the initial conditions which produce the 
observed properties (masses, \Teff\, radii, orbital period, etc.) of the current system by 
comparing them with calculated models, and finds the best match. This method is generally 
well suited to the analysis of a large sample of systems, and has been successfully applied 
by \citet{Nelson} and \citet{deMink}. The second approach calculates a series of models for 
a given system by guessing all possible initial configurations with angular momentum and mass 
loss assumptions and building binary evolution grids to find the best match. Nowadays this 
CPU-intensive method is possible thanks to modern computing power \citep{kolbas_uher}.

\begin{figure*}
\includegraphics[width=8cm]{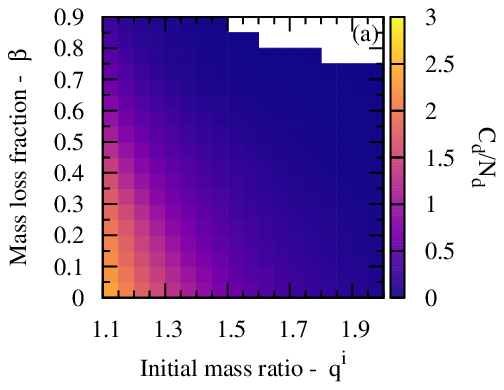}
\includegraphics[width=8cm]{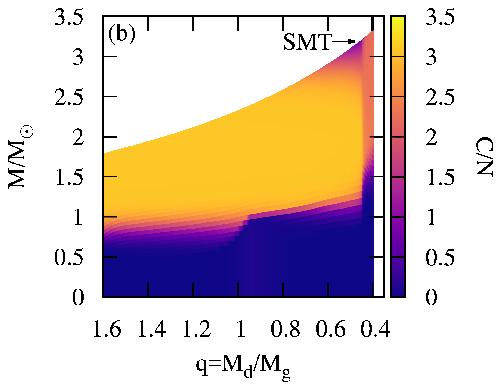}
\includegraphics[width=8cm]{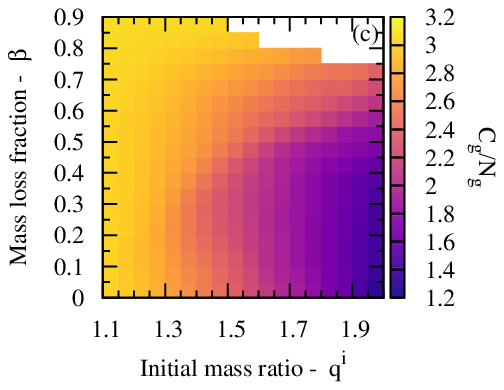}
\includegraphics[width=8cm]{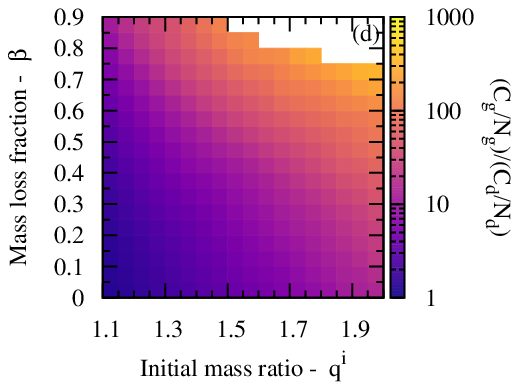}

\caption{The surface C/N yield of the mass-loser through grid is shown in 
panel (a) with associated colour scale. The interior C/N profile of the 
mass-gainer during the mass transfer, i.e.\ from higher mass ratio to 
lower one, is shown in panel (b). The end of rapid mass transfer (SMT) 
is annotated. The surface C/N yield of the mass-gainer through the grid 
is also shown in panel (c). While the individual outcome of the components' 
C/N ratio can be similar for a given $q^i$ and $\beta$, the ratio of the 
C/N ratios for the mass loser and gainer could be a good diagnostic to 
resolve the degeneracy in the system, as shown in panel (d).}

\label{figev}

\end{figure*}

The calculations of the evolutionary models were undertaken following the 
method and conventions presented in \citet{kolbas_uher}. The input ingredients
 of a model for a given metallicity are: the final masses of the components 
($M_{d}^{f}$, $M_{g}^{f}$), thus the final mass ratio 
($q^f=$ $M_{d}^{f}/M_{g}^{f}$), and the final orbital period ($P^f$). 
Then, the calculated quantities are the initial mass of the donor 
($M_{d}^{i}$), the initial mass of the gainer ($M_{g}^{i}$), and the 
initial orbital period ($P^i$) of the system for a given initial mass 
ratio ($q^i=$ $M_{d}^{i}/M_{g}^{i}$), and the mass transfer efficiency 
parameter ($\beta$). Using the assumption that mass loss 
(i.e.\ $(1-\beta)\,dM_d$) carries the angular momentum away from the donor 
\citep{Hurley, kolbas_uher}, a simple relation between the period, 
the total mass, and the components' masses is derived. 
Namely $P\,M_t^{2}\,M_g^{3}\,M_d^{3(1-\beta)} = {\rm const}$  during the orbital 
evolution.

Then the initial period of system can be derived from the equation
\begin{equation}
P^i=P^f    \left( \frac{M_t^f}{M_t^i} \right)^2 \left( \frac{M_g^f}{M_g^i} 
\right)^3 \left( \frac{M_d^f}{M_d^i} \right)^{3(1-\beta)} 
\label{fper}
\end{equation}
where
\begin{equation}
M_t^i=M_t^f\frac{(1+q^i)}{(1+q^f)}\frac{[1+q^f(1-\beta)]}{[1+q^i(1-\beta)]}.
\label{fmass}
\end{equation}
is the systemic mass equation as a function of initial and final mass 
ratio \citep{Giuricin}.

Equations (\ref{fper}) and (\ref{fmass}) also give the well-known equivalent 
relations for the assumptions of mass  (i.e.\ $\beta=0$) and  angular momentum 
conservations. As one can see, 
since the final parameters of the system 
(i.e., $M_d^f$,$M_g^f$, $M_t^f$, $q^f$, $P^f$) are known, the only unknowns 
are $q^i$ and $\beta$ which are taken to be free parameters.

Thus, using Eq.\ (1) and Eq.\ (2), one can calculate the possible initial 
parameters of a system (i.e., $M_d^i$, $M_g^i$, $M_t^i$, $P^i$) for various 
$q^i$ and $\beta$ values. These parameters can be used to build a grid of 
evolutionary tracks which allows us to compare, the current absolute 
parameters of the system ($M_{\rm A}$, $M_{\rm B}$, $T_{\rm eff,A}$, $T_{\rm eff,B}$, 
$R_{\rm A}$, $R_{\rm B}$) to find the best fitting models with accompanying $q^i$ and 
$\beta$.

Although the value of the efficiency parameters could be between 
$0 \le \beta \le 1$ (i.e.\ from conservative to totally non-conservative
 mass transfer), we found for $\delta$~Lib that some  of the initial periods 
with $\beta \gtrsim 0.8$ for a given $q^i$ are lower than the limiting period
 \citep{Nelson} for a given binary system, namely
\begin{equation}
P_{\rm lim}\approx\frac{0.19M_g^i+0.47{M_g^i}^{2.33}}{1+1.18{M_g^i}^2}.
\end{equation}
Thus we excluded all the initial parameters with $P^i \leq P_{\rm lim}$ 
from our analysis.

As also implied from Eq.\ (1), mass transfer leads to fast increase of
 the orbital period after mass ratio reversal. Thus $\delta$~Lib must 
have had a short initial orbital period to explain its current period. 
This situation constrains the number of initial parameters.

 For the initial mass ratio, we used the mean mass ratio of detached 
binaries from \citet{Ibanoglu} as a lower limit ($q^i \approx 1.1$). 
As a result of such consideration, we calculated the initial parameters 
of the system as $\beta=[0.0-0.9]$ with an increment size of 
$\delta\beta=0.05$ and $q^i=[1.1-2.0]$ with an increment size of 
$\delta q^i=0.05$ which constructs a 18$\times$18 matrix grid size. 
This is considerably higher resolution than those of \citet{kolbas_uher} 
which used a 4x5 grid size.

The Cambridge version of the 
{\sc stars}\footnote{{\tt http://www.ast.cam.ac.uk/\textasciitilde stars/}} 
evolution code was used for model calculations. {\sc stars} was originally 
developed by \cite{Eggleton1971,Eggleton1972} to calculate the evolutionary 
tracks for the initial parameters, $q^i$ and $\beta$. The  code was 
substantially upgraded for both new EOS tables and the capability of 
simultaneous calculation for both components of a binary system 
\citep{Pols, Stancliffe}. In our calculations, solar composition 
for both components is assumed, and the overshooting parameter is 
kept fixed at $\delta_{os}\sim0.12$.

We have run the code until the second mass transfer has occurred which 
is assumed to be the end of \textit{algolism} for binary systems. 
Considering that every initial configuration set produces approximately 
5000 interior models for each component, we have more than 3\,000\,000 
models of stellar interiors to compare to find the best fitting one.

For each initial set of masses and periods, we calculated the $\chi^{2}$ 
of the corresponding evolutionary tracks. Then we compared the minimum 
$\chi^2$ of every set of initial parameters with each other to find the 
best fitting model pair which matches both components with same age. 
We also calculated the likelihood of each minimum $\chi^2$ based on 
the derived errors of both components of the $\delta$~Lib system which 
are given in Table~\ref{tab:abspar}.

The components' physical parameters were derived from Fig.\,\ref{fighr} 
showing the HR diagram of the best fitting model tracks of both components 
together with the confidence countours from likelihood analysis. We plot 
the HR diagram of the best-fitting set of model tracks, confidence contours 
from the likelihood analysis and the derived physical parameters of the 
components in Fig.\,\ref{fighr}.


As a result of this, we determine the parameters of the best fitting model 
as $q^i=1.60$ and $\beta=0.0$. Using likelihood calculations, we determined 
the most probable initial parameters and associated errors as  
$\langle q^i \rangle = 1.626\pm0.125$ and $\langle\beta\rangle = 0.142\pm0.11$.
 Thus we determine the initial mass of the gainer and the period of the system 
as $M_d^i=3.028 \pm 0.161$ \Msun \,and $P^i=1.3531 \pm 0.099$ d.

This result is based on absolute parameters. From the abundance analyses we 
also know that C/N$ = 1.55\pm0.4$, which can be used for constraining the 
previous determination. So taking into account the thermohaline constraint, 
we limited the posible sets of initial conditions to those which satisfy 
the C/N ratio within an uncertainty range and determine the expected initial 
parameters $\langle q^i \rangle = 1.765\pm0.071$ and $\langle\beta\rangle = 
0.13\pm0.11$. This yields the initial mass of the gainer and the period of 
the system as $M_d^i=3.116 \pm 0.161$ \Msun \,and $P^i=1.435 \pm 0.108$ d. From both 
of these results we conclude that $\delta$~Lib has experienced a conservative 
to very mildly non-conservative evolution.

It is our principal interest to compare the chemical evolution outcome of 
the models with the results of the abundance determination for species that 
are most sensitive to CNO processing and mixing in stellar interiors prior 
to mass transfer. As was explained in the Introduction, C and N suffer the 
largest changes, and in some sense N becomes overabundant at the expense of 
C. For this purpose, we extracted the C/N ratios for each component from 
the minimum-$\chi^2$ models of every set of $q^i$ and $\beta$. In 
Fig.\,\ref{figev}a we show the outcomes for the mass donor's surface 
C/N ratio after mass transfer stripped the outer layers. The surface 
C/N ratio is lower than the solar value of (C/N)$_\odot \approx$ 2.8 
and decreases with increasing $q^i$ and $\beta$ down to C/N $\sim$ 0.01.

As we mentioned, the mass reversal leads to a period increase during 
the orbital evolution and thus ends the rapid phase of mass transfer. 
Since in non-conservative cases, a part of the transferred mass is not 
accreted, mass reversal occurs later than for conservative cases. 
This is a natural result of non-conservative orbital dynamics which 
leads to a postponed mass reversal and thus deeper layers are exposed.

As the mass ratio decreases in the course of the mass transfer, the deeper 
layers which contain nuclear processed material from the CNO cycle are 
exposed shortly after the mass ratio reversal has occurred. Based on the 
best fit parameters, it is expected that the surface C/N$_d=0.35\pm0.2$. 
Unfortunately, due to the small contribution of the donor to the total 
light of the $\delta$~Lib system, about 6 per cent (Table\,\ref{tab:lcsol}), 
the S/N of the donor's disentangled spectrum is not sufficient for a detailed
 abundance analysis with the required precision.

Once mass transfer begins, the $1-\beta$ fraction of the lost matter from 
the donor will be accreted by the gainer. As we have already shown, the 
accreted donor's material will have a different C/N ratio than that of 
the surface of the gainer which satisfies the starting condition of 
thermohaline mixing as described by \cite{Kippenhahn1980}. This leads to 
the mixing of the gainer's original (initial) material with the accreted 
one. According to the calculations, thermohaline mixing takes control just 
after the rapid mass transfer phase on a relatively short timescale of 
$10^5-10^6$ years (cf. \citep{kolbas_uher}).

In Fig.\,\ref{figev}b the internal C/N ratio profile of the mass gainer 
during mass transfer, i.e.\ decreasing the mass ratio, is shown. We have 
annotated the end of the rapid mass transfer phase, i.e.\ slow mass 
transfer (SMT), to show the efficiency of thermohaline mixing which 
smooths the abundance gradient of the outer layer in a short time.

The surface C/N ratio of the mass gainer from the minimum $\chi^2$ model 
through the grid is shown in Fig.\ 11c. If we compare Fig.\ 11a and Fig.\ 
11c, we see that while increasing $\beta$ for the donor leads more deeper 
layers to be exposed hence reduces the C/N ratio, the same increase leads 
to a smaller decrease in the C/N ratio of the gainer due to the fraction 
of matter not accreted. The decrease in C/N ratio with increasing $q^i$ 
remains similar for both components.

As a result of the mentioned confidence analysis, we determine 
the C/N ratio of the gainer from the models as C/N$_g = 1.79\pm0.14$. 
This has to be compared to  the value we determined from the abundance 
analyses in Sect.\,\ref{sec:abund} (C/N$ = 1.55\pm0.40$). The agreement 
between these two values is supporting the theoretical view, and expectation 
for almost conservative mass transfer in $\delta$~Lib.

Having two such different outcomes of the donor's and gainer's C/N ratio 
for an increasing $q^i$ and $\beta$ may lead us to another conclusion. 
In Fig.\,\ref{figev}d, we show the proportion of the C/N 
ratio of the mass donor and gainer. As one can see, the differences between 
the component's C/N more pronounced with increasing $q^i$ and $\beta$. While
 the ratio is around unity for a conservative and low initial mass ratio 
system, it goes up to 800 times that for very non-conservative and initially 
high mass ratio system. This also emphasises the importance of the abundance 
analysis on the faint component through spectral disentangling.

\section{Conclusion}

Mass and angular momentum transfer in Algol type binary systems, i.e.~in the 
first, and rapid phase in which mass reversal between the components happen, 
is still an open problem. As a consequence of this almost cataclysmic event, 
the photospheric chemical composition of the components could be altered. 
The C/N abundance ratio turns out to be a sensitive probe of the evolutionary 
processes, and in particular the thermohaline mixing (c.f.~\citet{Sarna}, 
and references therein). Tracing the C/N abundance ratio in the photospheric 
composition of the components could constrain their past, and eventually 
provide the initial stellar and system parameters. In the present work, 
we apply this concept to $\delta$~Lib, a classical Algol-type binary system.

In the following we summarise and conclude our study:

\begin{enumerate}

\item New ground-based high-resolution and high-S/N \'echelle spectroscopy 
was secured. A new series of observed spectra were used for determination 
of the orbital elements. The RVs are affected by the RM effect in the course 
of the eclipses, tidal distortion, and a cool obscuring matter between the 
components, particularly for the mass-losing component. The measurements 
obtained with cross-correlation were used to account properly for these 
non-Keplerian effects. Only when the influence of dark material in the 
form of a cool spot on star B was taken into consideration, were 
satisfactory fits for the RVs variation obtained. This reduced 
significantly the masses of the components, and in particular 
the mass of star A (Table\,\ref{tab:orbit}). In turn, the mass 
of star A is now more compatible with its radius and \Teff, 
a long-standing issue for the $\delta$~Lib system. The case of 
a dark, cool spot distorting RVs for the mass-losing component 
in RZ~Cas was previously discussed by \citet{Tkachenko_rzcas}.

\item Separated spectra of the components obtained using the method 
of spectral disentangling was used for the determination of the 
atmospheric parameters (Table\,\ref{tab:abspar}).

\item The LC from the predominantly $R$ passband from the {\sc stereo} 
mission space photometry of $\delta$~Lib was analysed, along with LCs 
in the optical and near-IR compiled from previous studies. Complementary 
RV and LC analysis yielded the physical properties for both components 
with an improved precision compared to previous studies 
(Table\,\ref{tab:abspar}: 3.5 percent for star A's mass and 
radius, 2.7 percent for star B's mass, and 1 percent for star B's 
radius. The uncertainty in the radius of the Roche-lobe filling star 
B depends solely on the precision in the mass ratio, so is considerably 
smaller than for star A.

\item The photospheric chemical composition was determined for star A. 
We found a metallicity [Fe/H] $= 0.15 \pm 0.09$, which is higher than 
the solar value. This supports a higher abundances of metals compared 
to solar values, too. Its C/N ratio is $1.55\pm0.40$, also altered since 
the solar value is (C/N)$_\odot = 2.80\pm0.30$ \citep{asplund}.

\item Almost 3 million structure models were calculated with the Cambridge 
{\sc stars} evolutionary code to match the observed properties of the 
components in $\delta$~Lib after a mass reversal. These models were also 
parametrised with $\beta$, an indicator of mass-loss efficiency. 
Constraining the initial models with the observed gainer's C/N abundance 
ratio, we found that mass transfer was either mildy non-conservative or 
fully conservative, with $\beta = 0.14\pm0.11$.

\end{enumerate}

The evolutionary modelling for Algol-type systems performed in the present 
work would be substantially constrained by including the C/N abundance ratio 
for the secondary component, too. Whilst this means a significant increase 
in observational effort to gain sufficiently high S/N for a faint secondary 
component, it could be rewarding by giving a more accurate determination of 
the secondary's RV curve. Due to tidal distortion of the subgiant shape, 
its RV curve is also affected. This is an important effect in Algol-type 
binaries, and could affect the RV semi-amplitude of the distorted component 
by several\kms, as was first found by \citet{Andersen} following the 
theoretical prediction of \citet{Wilson79}. However, it was found that 
more influential on the secondary's RV variations, is an obscuring dark 
cloud of material between the components, also revealed in Doppler 
tomography \citep{Richards_tomo}. Taking into account non-Keplerian 
effects on the RV variations, the determined mass of star A was found 
to be more consistent with its radius and \Teff.

Extensive spectroscopy of $\delta$~Lib during eclipse could also resolve 
the discrepancy between the observed equatorial velocity for star A and 
its calculated synchronous velocity. The RM effect is pronounced in the 
primary eclipse, and could be used for an independent determination 
of  its rotational velocity. 
And last, but not least, an improvement in the precision 
of the radii is needed. Our analysis of available {\sc stereo} space 
photometry is an improvement in this direction, but was not sufficient. 
Space based, all-sky, bright star photometry like the {\sc tess} 
\citep{Ricker_tess} mission would be indispensable in that sense.

\section*{Acknowledgements}

Observations were collected at the Centro Astron\'omico Hispano Alem\'an 
(CAHA) at Calar Alto are operated jointly by the Max-Planck Institut 
f\"ur Astronomie and the Instituto de Astrof\'{\i}sica de Andaluc\'{i}a 
(CSIC), and Th\"{u}ringer Landessternwarte Tautenburg, Germany.

KP is financially supported by the Croatian Science Foundation through 
grant IP 2014-09-8656, which also enables a postdoctortal fellowship 
to AD. AD also acknowledge support by the Turkish Scientific and 
Technical Research Council (T\"{U}B\.{I}TAK) under research grant 113F067.

The authors would like to thank Karl Wraight (formerly of the Open University)
 for the preliminary analysis of the {\sc{stereo}} photometry of $\delta$\,Lib.
 We thank the anonymous referee for the constructive comments.

\bibliographystyle{mnras}

\label{lastpage}

\end{document}